\documentclass[apj,twocolumn]{openjournal}
\usepackage{amsmath}
\usepackage[usestackEOL]{stackengine}
\setstackgap{S}{-0.5pt}
\setstackgap{L}{6.5pt}
\usepackage{booktabs,comment}

\usepackage[dvipsnames]{xcolor} %added later for color
\usepackage[breaklinks,colorlinks,citecolor=blue,urlcolor=blue]{hyperref}

\usepackage{orcidlink}
\usepackage{subfigure,comment}

\renewcommand{\arraystretch}{1.2}  
\usepackage{listings}
\usepackage{color}
\definecolor{dkgreen}{rgb}{0,0.6,0}
\definecolor{gray}{rgb}{0.5,0.5,0.5}
\definecolor{mauve}{rgb}{0.58,0,0.82}
\definecolor{golden}{rgb}{0.86,0.65,0.01}
\usepackage{soul}
\usepackage{amsmath}
\usepackage{xspace}
\usepackage{graphicx}
\usepackage{enumitem}
\usepackage{amssymb}
\usepackage{xifthen}
\usepackage{hyperref}
\usepackage[normalem]{ulem}
\usepackage{multirow}
\newcommand{\code}[1]{\texttt{#1}\xspace}

\newcommand{\unit}[1]{\ensuremath{\mathrm{\,#1}}\xspace}

\newcommand{\kms}{\unit{km\,s^{-1}}}

\newcommand{\msun}{\unit{M_\odot}}
\newcommand{\lsun}{\unit{L_\odot}}
\newcommand{\LCDM}{$\Lambda$CDM\xspace}

\begin{document}

% \preprint{}

\author{Pierre Thibodeaux\,\orcidlink{0000-0002-3867-3927}$^{1,2}$}
\author{Alex Drlica-Wagner\,\orcidlink{0000-0001-8251-933X}$^{3,1,2,4}$}
\author{Alexander P. Ji\,\orcidlink{0000-0002-4863-8842}$^{1,2,4}$}
\author{Chin Yi Tan\,\orcidlink{0000-0003-0478-0473}$^{5,2}$}
\author{Ting S.~Li\orcidlink{0000-0002-9110-6163}$^{6}$}
\author{Andrew~B.~Pace\,\orcidlink{0000-0002-6021-8760}$^{7}$}
\author{Louis E. Strigari\orcidlink{0000-0001-5672-6079}$^{8}$}

\affiliation{$^1$Department of Astronomy \& Astrophysics, University of Chicago, 5640 S Ellis Avenue, Chicago, IL 60637, USA}
\affiliation{$^2$Kavli Institute for Cosmological Physics, University of Chicago, Chicago, IL 60637, USA}
\affiliation{$^3$Fermi National Accelerator Laboratory, P.O. Box 500, Batavia, IL 60510, USA}
\affiliation{$^{4}$NSF-Simons AI Institute for the Sky (SkAI),172 E. Chestnut St., Chicago, IL 60611, USA}
\affiliation{$^{5}$Department of Physics, University of Chicago, 5720 S Ellis Avenue, Chicago, IL 60637, USA}
\affiliation{$^{6}$ Department of Astronomy and Astrophysics, University of Toronto, 50 St. George Street, Toronto ON, M5S 3H4, Canada}
\affiliation{$^{7}$ Department of Astronomy, University of Virginia, 530 McCormick Road, Charlottesville, VA 22904 USA}
\affiliation{$^{8}$ Department of Physics and Astronomy, Mitchell Institute for Fundamental Physics and Astronomy, Texas A\&M University, College
Station, Texas 77843, USA}

\begin{abstract}

Ultra-faint dwarf galaxies (UFDs)---the smallest and most dark-matter dominated galactic systems---push the boundaries of our understanding of galaxy formation in the cold dark matter paradigm. As current and near-future imaging surveys continue to expand the known population of extremely faint systems, spectroscopic resources have become a major limitation in our ability to understand the physical nature of these systems. We present projections for the capability of Keck/DEIMOS, Magellan/IMACS, and future spectroscopic instruments (e.g., Magellan/Via and GMT/GMACS) to kinematically distinguish dark-matter-dominated UFDs from dark-matter-free star clusters in the ultra-faint regime. We simulate ensembles of mock observations of ultra-faint dwarf galaxies and apply a formal statistical framework, calibrated to observations, to systematically characterize the ability to resolve velocity dispersions as a function of satellite properties. Our projections suggest that current instruments will struggle to resolve many of systems that will be discovered by the Rubin Observatory. Near future instruments will be capable of resolving a majority of these systems, though they will fail for the faintest and most distant systems. Ultimately, we will require even more advanced instruments, with enhanced stability and velocity precision, to classify the full population of ultra-faint Milky Way satellites.

\end{abstract}

\title{Forecasting the Resolvability of Stellar Velocity Dispersions in Ultra-faint Dwarf Galaxies}

\maketitle

\section{Introduction}
\label{sec:intro}
The standard cosmological model consisting of a cosmological constant and cold dark matter (\LCDM) does an excellent job reproducing many observables, such as the Cosmic Microwave Background, Big Bang nucleosynthesis, galaxy clustering, weak lensing, and the Lyman-$\alpha$ forest \citep{peebles_large-scale_1982,blumenthal_formation_1984, chabanier_matter_2019,the_planck_collaboration_planck_2020}. 
While \LCDM has had great success modeling large-scale structure, it initially faced challenges in modeling smaller scales. Apparent tensions between \LCDM simulations and observations demonstrated that Milky Way dwarf galaxies are promising probes of novel dark matter physics \citep{bullock_small-scale_2017,primack_galaxy_2024}. 

The smallest observed galaxies, ultra-faint dwarfs (UFDs), allow us to study how ancient galaxies formed and evolved \citep{simon_faintest_2019}. The star formation histories of UFDs were affected by reionization, making them a powerful probe of when and how reionization occurred \citep[e.g.,][]{bose_imprint_2018}. As the most chemically pristine systems, they are excellent places to find metal-poor stars and stars with extreme compositions, which provide insight into stellar evolution and how those stars influenced the early universe \citep[e.g.,][]{magg_predicting_2018, ji_spectacular_2024}.

UFD candidates are first identified as stellar overdensities in photometric surveys, requiring spectroscopic followup to better characterize their physical properties and distinguish them from star clusters. Applying the definitions from \citet{willman_galaxy_2012}, a dwarf galaxy resides in a dark matter halo while star clusters are self-gravitating baryonic systems. Determining whether a UFD candidate is a galaxy comes down to searching for observable signatures that imply the presence of dark matter. Stellar velocity dispersions are the gold standard for UFD classification: in a virialized system, a large velocity dispersion is associated with a deep gravitational potential. When the depth of this potential far exceeds predictions from visible baryonic matter alone, it is attributed to the presence of dark matter. 

 Massive star clusters, $M_\star\sim 10^5\,M_\odot$, can have substantial self-gravitating velocity dispersions, but similarly bright UFDs have much larger dispersions (for example, Eridanus II has a reported velocity dispersion of 6.9\,\kms, but would have a self-gravitating velocity dispersion of $\sim 0.4$\,\kms; \citealp{li_farthest_2017}, using estimator from \citealp{errani_ursa_2024}). Smaller star clusters, $M_\star <10^3 M_\odot$, are expected to have self-gravitating dispersions $<0.1$\,\kms, whereas similarly bright UFDs still have velocity dispersions on the scale of a few \kms (\citealp{wolf_accurate_2010}; see also \citealp{smith_discovery_2024}). For the smallest candidates, we say that a system's velocity dispersion is \emph{resolved} if it can be shown that the dispersion is statistically inconsistent with zero; a UFD candidate with a resolved dispersion is usually classified as a dwarf galaxy.

Many UFD candidates have been found in current imaging surveys (e.g., DELVE, \citealp{drlica-wagner_decam_2021}; UNIONS, \citealp{gwyn_unions_2025}). We expect to discover many more in upcoming surveys (e.g., Euclid, \citealp{euclid_collaboration_euclid_2025}; LSST, \citealp{ivezic_lsst_2019}; Roman HLWAS, \citealp{observations_time_allocation_committee_roman_2025}). However, these candidates will be smaller and fainter than previously observed systems, making it more difficult than ever to successfully resolve their velocity dispersions \citep{tsiane_predictions_2025}. Since these systems will have few stars bright enough to access with spectroscopy, measurement of their dispersions will extremely sensitive to the choice of a prior, which varies from study to study \citep[e.g.,][]{fritz_gaia_2019,li_ships_2018}. Moreover, the criteria to evaluate whether a system is \emph{resolved} differs between analyses---as an example, we point to Grus II \citep{simon_birds_2020} and Tucana V \citep{hansen_chemical_2024} which are reported with the same dispersions and similar uncertainties, but with different conclusions. Fainter systems with larger velocity uncertainties will be even more ambiguous.

These issues motivate us to standardize the process of claiming a resolved velocity dispersion in a candidate UFD with a small sample. Having a consistent definition allows us to generate projections for our ability to resolve UFDs of varying properties with several instruments. This will also enhance ensemble modeling of UFD velocity dispersions \citep[e.g.,][]{esteban_milky_2024} by accounting for regions of parameter space where we expect UFD dispersions to be inaccessible.

In this paper, we perform a systematic study to characterize resolvability limits for current and near-future instruments. In Section \ref{sec:theory}, we introduce a literature-calibrated hypothesis test formalism to determine whether a dwarf’s velocity dispersion is resolved. In Section \ref{sec:application}, we describe the application of this test to literature velocities from UFD candidates. In Section \ref{sec:methods}, we discuss our procedure for simulating mock observations of UFD stellar velocities. In Section \ref{sec:projections}, we make projections with DEIMOS and IMACS to guide the planning of spectroscopic followup observations of UFD candidates. We also explore the implications of our projections on the population of UFD candidates that is expected to be discovered by the Rubin Observatory's Legacy Survey of Space and Time (LSST), and we consider how future instruments, such as Viaspec and GMACS, will attempt to resolve this population. Finally, in Section \ref{sec:disc} we discuss caveats of our analysis, concluding in Section \ref{sec:conclusion}.

\section{Statistical Formalism}
\label{sec:theory}
The line-of-sight velocity dispersion for a Milky Way satellite is commonly derived from a sample of member stars using the likelihood procedure described in \citet{walker_internal_2006}. The radial velocities, $v_i$, and uncertainties, $e_i$, of $N$ sample stars are passed to a Gaussian likelihood function, 
%Gaussian-distributed data with heteroscedastic, Gaussian errors:
\begin{align}
\label{eqn:lh}
    \mathcal{L}(v_i,e_i |\sigma_v,\mu_v ) =\prod_{i=1}^{N} \frac{1}{\sqrt{2\pi (\sigma_v^2+e_i^2)}}\exp \left(-\frac{(v_i-\mu_v)^2}{2(\sigma_v^2+e_i^2)}\right),
\end{align}
where $\sigma_v$ and $\mu_v$ are the intrinsic velocity dispersion and mean (systemic) velocity respectively. Then, most analyses use Markov Chain Monte Carlo (MCMC) to obtain the posterior distributions of $\mu_v$ and $\sigma_v$, from which they quote a velocity dispersion and its uncertainty.

This method provides no unique criterion for declaring when a velocity dispersion is sufficiently separated from zero to claim that a system is resolved. In reality, authors make various choices about the procedure and threshold that should be used when claiming the detection of a resolved velocity dispersion. Here, we define a formal hypothesis test that allows a consistent analysis to be performed across data and simulations. 

%In order to make projections from simulations, we need a definition that can be decided without unique intervention. It is additionally desirable that this could standardize UFD followup analyses. To solve these issues, we turn to hypothesis testing.

\subsection{Hypothesis Testing}

Hypothesis testing involves calculating a statistic from a sample of member star velocities and testing it against a threshold in order to decide whether a system has a resolved velocity dispersion. This is a common statistics problem where we seek to choose from two nested hypotheses, one of which resides on the boundary of the allowed parameter range. The null hypothesis, $H_\mathrm{null}$, is that the system is unresolved with $\sigma_v=0$. The alternative hypothesis, $H_\mathrm{alt}$, is that a system is resolved with a nonzero velocity dispersion, $\sigma_v$.  Our goal is to determine the statistical preference (or lack thereof) for the alternative hypothesis relative to the null hypothesis. 

% $\sigma_v=0$ is a specific case of $\sigma_v\geq0$, and $\mathbb{P}(\sigma_v\geq 0)=\mathbb{P}(\sigma_v>0)$.

%evidence for either hypothesis, and if it evidences the alternative hypothesis by a sufficient margin, we reject the null hypothesis in its favor.

%Commonly, hypothesis testing is eschewed for direct parameter inference, because the latter provides a numerical answer instead of a binary one. However, in this situation, we actually care about the binary answer of this hypothesis test, since it influences how we handle the analysis, whether we report a resolved dispersion or an upper limit.

%The key is that we consider how strongly the data evidences \textit{both} hypotheses. We expect that most choices that one could make during parameter inference (method of fitting, numerical choices, choice of priors) would affect both measures of evidence in the same way, canceling out their effects. 

\begin{figure}
    \centering
    \includegraphics[width=\linewidth]{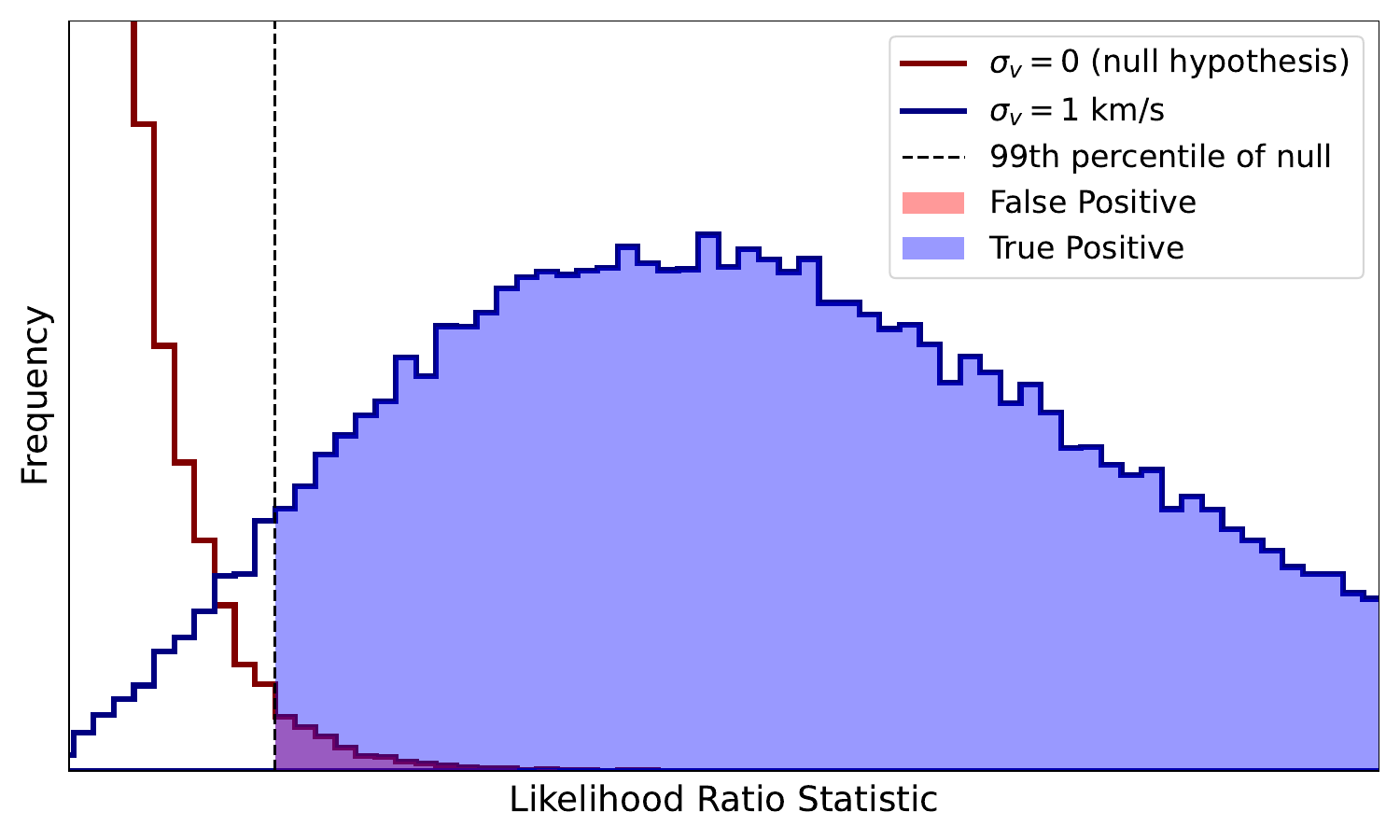}
    \caption{In this schematic, the left (red) distribution describes the likelihood ratio statistic $\lambda_\mathrm{lh}$ for instances following the null hypothesis, and the right (blue) distribution describes the LHRs for instances following an alternative hypothesis. Applying a threshold value for claiming detection of a resolved velocity dispersion (dotted line) admits some fraction of \emph{false positives}, represented by the shaded red region. Conversely, the efficiency of identifying \emph{true positives} is shown by the blue region.}
    \label{fig:falsepos}
\end{figure}

To evaluate this comparative evidence, we define the likelihood ratio test statistic, $\lambda_\mathrm{lh}$ (specifically the Wilks' test statistic),
\begin{align}
    \lambda_\mathrm{lh}&=2\ln\left(\frac{\max[\mathbb{P}(\mathrm{data} |H_\mathrm{alt})]}{\max[\mathbb{P}(\mathrm{data} |H_\mathrm{null})]}\right)\\
    \notag &=2\ln\left(\frac{\max[\mathcal{L}(\mathrm{data} |\sigma_v)]}{\max[\mathcal{L}(\mathrm{data} |\sigma_v=0)]}\right).
\end{align}
 $\mathcal{L}$ is maximized by the maximum likelihood estimator (MLE). Thus we can write,
\begin{align}
    \label{eqn:LHR}
    \lambda_\mathrm{lh} &=2\ln \left(\frac{\mathcal{L}(\mathrm{data}|\sigma_v^\mathrm{MLE})}{\mathcal{L}(\mathrm{data}|\sigma_v=0)}\right).
\end{align}
This likelihood ratio (LHR) is a frequentist statistic that compares how likely a dataset is to have come from either of the two hypotheses. In order to favor the alternative over the null hypothesis, this statistic needs to exceed some threshold, based on how often random draws from the null hypothesis distribution appear to violate the null hypothesis. We depict this visually in Figure \ref{fig:falsepos}, showing how we base the LHR threshold on a false positive rate, resulting in some true positive efficiency. The completeness of true positive UFD identifications is naturally in tension with the number of false positives we admit.

Under a set of well defined asymptotic conditions, this threshold can determined analytically from Chernoff's extension of Wilks' theorem \citep{wilks_large-sample_1938,chernoff_distribution_1954}, which states that the LHR of the null hypothesis is distributed as half $\chi^2_1$ distribution and half delta function centered on 0.\footnote{Wilks' theorem says that $\lambda_\mathrm{lh}$ should be distributed as $\chi_k^2$ for $k$ variables, and Chernoff's extension applies in the case where $k=1$ and the variable lies on the boundary of the parameter space, such as in our case, $\sigma_v>0$.} However, this analytic description only applies exactly in the asymptotic limit of large samples, which may not be true for UFD observations. Instead, in Appendix \ref{app:deri} we derive an analytic description of the distribution of $\lambda_\mathrm{lh}$ directly from our likelihood function and ratio (Equations \ref{eqn:lh} and \ref{eqn:LHR}). This results in a threshold of
\begin{equation}
\label{eqn:thresh}
    \Lambda_\mathrm{th}=X_\alpha-N-N\log(X_\alpha/N),
\end{equation} where $X_\alpha$ is the $1-\alpha$ percentile of $\chi^2$ with degrees of freedom $N-1$. In the limit of large N, this converges to Chernoff's extension. In deriving this equation, we assumed that: (1) the velocity data is sampled from a Gaussian distribution, and (2) the velocity uncertainties are homoskedastic. While these assumptions do not strictly hold for observations of Milky Way satellites, in Appendix \ref{app:deri} we demonstrate that the distribution of null-case LHRs is consistent with our derived distribution over a large range of simulated parameters.

In our analysis, we assumed a zero-dispersion case as the null hypothesis, even though the stellar mass of a dark-matter-free system would still imply a non-zero dispersion. However, for the faintest candidates, this distinction is minor; for example, the self-gravitating dispersion for a system like Ursa Major III/UNIONS I ($M_\star\sim16\,M_\odot$) is around $\sigma_\star\sim 0.05\,$\kms, far below our systematic velocity uncertainty \citep{errani_ursa_2024}. We note that this statistical framing does not account for systematic effects that could inflate the velocity dispersion (i.e., unidentified binaries; Section \ref{sec:binaries}).

\subsection{Significance Level}
In this analysis, we define the significance level to be the same as the false positive rate, where an $\alpha\%$ significance implies that only $\alpha\%$ of null cases have a test statistic, $\lambda_\mathrm{lh}$, that is larger than the threshold value. This is depicted in Figure \ref{fig:falsepos}, where it can be seen that completeness and purity are naturally in tension. The significance level required for any particular population-level analysis depends on the individual requirements on completeness and purity.

%For example, recent paper, \citet{may_updated_2025}, explains how the smallest UFDs could be used to constrain the mass of fuzzy dark matter, and one would desire a particularly pure sample of dwarf galaxies and therefore a particularly small significance level. 

\begin{figure}
    \centering
    \includegraphics[width=1.05\linewidth]{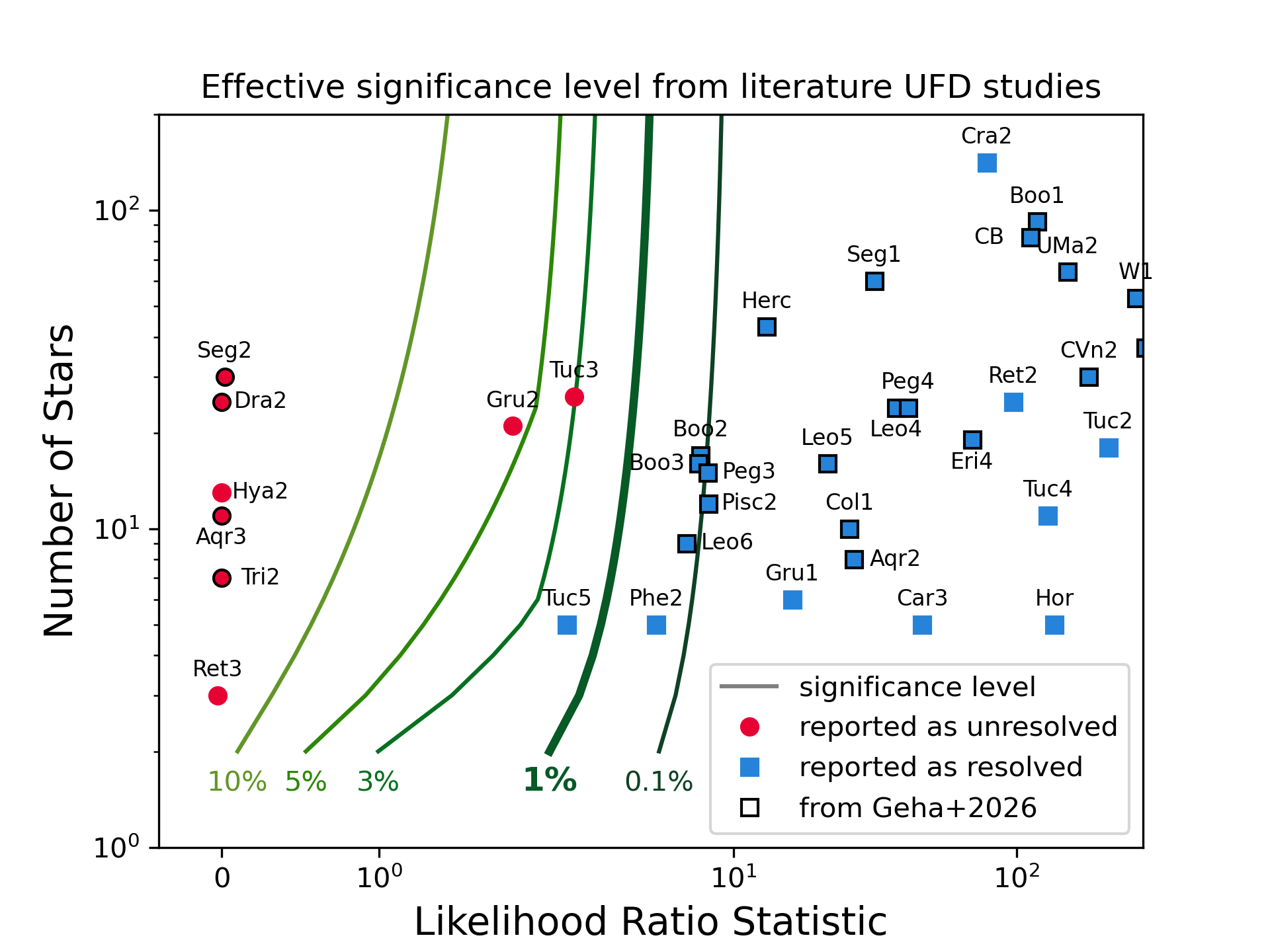}
    \caption{The likelihood ratio from comparing the null hypothesis ($\sigma_v = 0$\,\kms) to the alternative hypothesis ($\sigma_v > 0$) for the systems listed in Table~\ref{tab:ufds}. Systems that have been reported as ``resolved'' in the literature are shown with blue squares, while systems reported as ``unresolved'' in the literature are shown with red circles. Also shown are the significance threshold curves, $\Lambda_\mathrm{th}$,  derived from Equation \ref{eqn:thresh}. We find that a significance of $1\%$ matches the community's standard for declaring a resolved velocity dispersion despite the diverse set of criteria used in the literature.}
    \label{fig:effective_thresh}
\end{figure}

Rather than arguing for a specific threshold value,  we attempt to formalize the significance threshold that has been implicitly adopted in the literature. To determine this threshold, we calculate the LHR test statistics of an ensemble of Milky Way UFDs described in Section \ref{sec:application} and compiled in Table \ref{tab:ufds} in Appendix \ref{app:ufd}. We plot these $\lambda_\mathrm{lh}$ in Figure \ref{fig:effective_thresh}. The points are colored by whether they are considered resolved in their original analyses. Below a certain LHR, systems are reported as unresolved. We overplot the threshold curves from Equation \ref{eqn:thresh} and find that a significance level of $\sim$1--3\% broadly agrees with the literature. Here, we adopt a significance level of 1\%, which agrees with literature classifications with the exception of Tucana~V. Tucana~V was most recently reported as \emph{marginally resolved} \citep{hansen_chemical_2024}, but they also reported an upper limit. 

We note that the Bayes factor classification procedure described in \citet{cerny_chemodynamical_2026} is functionally a Bayesian implementation of the procedure we describe. However, rather than interpreting their threshold with a false positive rate, their method relies on a qualitative scale from \citet{kass_bayes_1995} to contextualize support for the alternative hypothesis. \citet{cerny_chemodynamical_2026} offer that a Bayes factor of $2\ln\beta >2$ suggests a \textit{weak preference} for a nonzero dispersion, $2\ln\beta >6$ is \textit{strong evidence}, and $2\ln\beta >10$ is \textit{very strong evidence}. For a system with 5 measured-stars, like the smallest in their census, these cutoffs correspond to significances of 2.8\%, 0.2\%, and 0.02\%. Figure~\ref{fig:effective_thresh} shows that the \textit{weak preference} (i.e., $2\ln\beta >2$) from \citet{cerny_chemodynamical_2026} is consistent with the community consensus for classifying UFDs.

\section{Application to Real Systems}
\label{sec:application}
As a systematic demonstration, we consistently apply our hypothesis test procedure to previously reported UFDs. This procedure consists of:
\begin{enumerate}[nosep]
    \item [1)] Calculating the LHR test statistic for the stellar sample using Eqs.~\ref{eqn:lh} and \ref{eqn:LHR}.
    \item [2)] Comparing the test statistic to the threshold value, from Eq. \ref{eqn:thresh}, at the adopted significance level of 1\%.
    \item [3a)] If the test statistic is greater than the threshold value, the system passes the hypothesis test and is considered resolved/informative. We report the best-fit value and the uncertainties.
    \item [3b)] If the test statistic is less than the threshold value, the system fails the hypothesis test. We instead report an upper limit.
\end{enumerate}

In Table \ref{tab:ufds} of Appendix A, we list each UFD system for which we have spectroscopic data, as well as the instrument used to observe it, the number of stars with measured velocities, and the studies that published those velocities. These were sourced from various analyses, including the recent archival DEIMOS study by \citet{geha_keckdeimos_2026-1}, which re-reduced data for many UFDs, reporting different stellar velocities, memberships, and final velocity dispersions. As noted in the Local Volume Data Base \citep[LVDB;][]{pace_local_2025}, these systems span luminosities of $10^{2.5}$ to $10^9\,L_\odot$ and distances of 21 to 370\,kpc, with sample sizes as small as 3 stars. 

It is only possible to identify binaries with multiple epochs of data, which is not always feasible for systems on the boundaries of observability. In our dataset, stars with velocity variability have had their velocity uncertainties increased, and binaries were removed when identified. Especially for the smallest UFDs, binaries can inflate measured dispersions, adding a few \kms \citep{mcconnachie_revisiting_2010}. We discuss complications and potential directions related to binary stars in more detail in Section \ref{sec:binaries}.

The threshold LHR, $\Lambda_\mathrm{th}$, (from Equation~\ref{eqn:thresh}), the measured test statistic, $\lambda_{\rm LH}$, and the result of our hypothesis test are displayed in Table \ref{tab:ufds}, reflecting their position in Figure \ref{fig:effective_thresh}. The dispersions are derived by sampling Equation \ref{eqn:lh} using an MCMC implemented in \code{emcee} \citep{foreman-mackey_emcee_2013}. We apply a uniform prior for $\sigma_v$ from 0.001\kms to 20\kms. We use 20 walkers with 500 steps, and a burn-in period of 100 steps, culling samples resulting in a negative dispersion. We generate a posterior distribution of $\sigma_v$, marginalizing over $\mu_v$, with a Gaussian kernel density estimator and report the dispersion as the maximum \emph{a posteriori} value. The uncertainties are derived from the maximum density interval that contains 68\% of the posterior. When we report an upper limit, it contains 90\% of the posterior.

\section{Simulating Mock Observations}
\label{sec:methods}
To characterize the range of parameters where systems will be resolvable, we simulate ensembles of observations of dwarf galaxies in halos with varying properties. Our code \code{dsphsim}\footnote{\url{https://github.com/pierrethx/dsphsim/}} draws from two sources: \code{StarSampler}\footnote{\url{https://github.com/maoshenl/StarSampler}} to generate velocities from an underlying gravitational potential and \code{ugali} \citep{bechtol_eight_2015}\footnote{\url{https://github.com/DarkEnergySurvey/ugali}}  to generate stellar magnitudes from an isochrone. \code{dsphsim} then generates uncertainties on those velocities based on observational data. 

\subsection{Dwarf Galaxy Scaling Relations}
\label{subsec:scaling}
We assume that the systems are spherical, in dynamic equilibrium, and have an isotropic velocity dispersion. We do not include tidal disruption for our systems. We use a Plummer profile \citep{plummer_problem_1911} for the stellar distribution and an NFW profile \citep{navarro_universal_1997} for the DM halo. 

\begin{figure*}
    \centering
    \includegraphics[width=\linewidth]{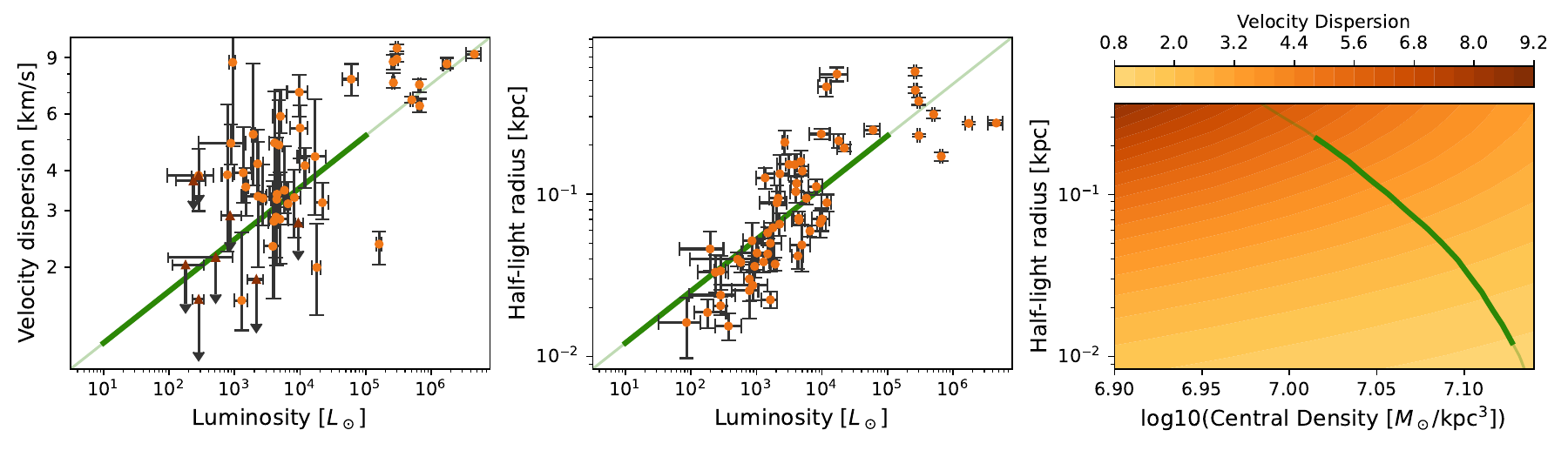}
    \caption{UFD scaling relations for the velocity dispersion (left) and half-light radius (middle) as a function of luminosity. In the left plot, resolved dispersions are represented with orange circles and upper limits are represented with brown triangles. These relations are combined with a concentration parameter from \citet{correa_accretion_2015} to derive central densities that agree with the velocity dispersion scaling relation (right). The green line in the right plot shows the scaling relation of central density with half-light radius, which relates central density and scale radius to luminosity.}
    \label{fig:scalingrel}
\end{figure*}

We then use the galaxies in the Local Volume Database (LVDB, \citealp{pace_local_2025}) to derive empirical scaling relationships to generate mock dwarfs as a function of one variable, $V$-band luminosity. We plot the velocity dispersions from Table \ref{tab:ufds} and half-light radii from the LVDB as functions of the luminosity in the first two panels of Figure \ref{fig:scalingrel}. Our simulated galaxies are assumed to follow the power-law relationship in green:
\begin{align}
    r_h &= 0.0058\, (L_\mathrm{V}/\mathrm{L}_\odot)^{0.319}\,\mathrm{kpc}\\
    \sigma_v &= 0.8\, (L_\mathrm{V}/\mathrm{L}_\odot)^{0.162}\,\mathrm{km\,s}^{-1}
\end{align}\noindent These relations are similar to those derived for DEIMOS satellites by \citep{geha_keckdeimos_2026}.

To relate the dark matter halo parameters to the luminosity, we introduce the analytic concentration--mass relation from \citet{correa_accretion_2015}. This fixes the halo mass and scale radius at a given central density. We then use \code{StarSampler} generate velocity dispersions for systems on a grid in steps of central density and half-light radius. We interpolate the value of central density that makes the velocity dispersions and half-light radii match their scaling relations. The grid and the central density scaling are plotted in the rightmost panel of Figure \ref{fig:scalingrel}.

Therefore, the simulated UFDs are completely specified by their luminosity. At the faintest end that we simulate for this work, $10$\lsun, the velocity dispersion is $\sim 1$\,\kms, the half-light radius is $\sim 12\,\mathrm{pc}$, the halo mass within the half-light radius is $\sim1.6\times10^4\,$\msun, which is around 20 times smaller than that of Segue 1 \citep{bissonette_metallicity_2026}. 

\subsection{Stellar Population Generation}
\label{subsec:ugali}

We adopt a stellar-mass-to-luminosity ratio of M$_\star$/$L_\mathrm{V}=2$, which is typical for faint dwarf galaxy satellites. We use \code{ugali} to convert the stellar mass into a number of stars for the system. We adopt a \citet{chabrier_galactic_2003} initial mass function and a 12-Gyr, $Z = 0.0002$ isochrone from \citet{bressan_parsec_2012} to generate DES $g$- and $i$-band magnitudes for the stars. At this stage, we remove horizontal branch (HB) stars; HB stars are commonly identified in more luminous systems, but hotter stars tend to have larger velocity uncertainties \citep[e.g.,][]{kirby_multi-element_2010,verdiani_carbon_2026}. 

We then generate positions and velocities of stars using \code{StarSampler}. These values are generated independently from the stars' masses. \code{StarSampler} produces 6-dimensional (position and velocity) samples of stellar systems from user-specified spatial distributions and gravitational potentials. Although the code allows for the direct sampling of the 6-d phase space through importance or rejection sampling, we use conditional sampling because of its high efficiency for generating large samples. Essentially, it first samples radii from the given spatial distribution and then samples velocities from the velocity distribution at that radius. We record the $z$-component of the velocity as the line-of-sight velocity for the stars, ignoring the effects of perspective motion \citep{walker_systemic_2008,kaplinghat_proper_2008}. We do not truncate our simulated galaxies at any radii. 

\subsection{Stellar Velocity Uncertainties}
We then simulate realistic observational uncertainties for our systems. For each star, the generated magnitudes are converted to signal-to-noise ratios (SNRs) based on data for the instrument and the given observation time. Our IMACS magnitude-to-SNR curve is based on an observation of Eridanus II from \citet{li_farthest_2017}. Our DEIMOS magnitude-to-SNR is based on the DEIMOS observation calculator.\footnote{\url{https://etc.ucolick.org/web_s2n/deimos}} Stars with ${\rm SNR} < 5$ are considered inaccessible for radial velocity measurements, serving as a lower magnitude limit. SNR is converted to statistical velocity measurement uncertainty based on \citet{simon_kinematics_2007}\footnote{This relation is reexamined in \citet{geha_keckdeimos_2026-1} and remains qualitatively similar, so we continue to use the original relation.} for DEIMOS and \citet{heiger_reading_2024} for IMACS. This statistical uncertainty is combined in quadrature with a systematic velocity uncertainty for the instrument to assign an estimate of the total velocity uncertainty. In this work, we assume a systematic velocity uncertainty of 1.1\,\kms for both DEIMOS \citep{geha_keckdeimos_2026-1} and IMACS \citep{heiger_reading_2024}. Lastly, the observed velocity for each star is randomized from its true velocity by drawing a random offset from a normal distribution with a standard deviation set by the quadrature sum of the statistical and systematic uncertainty.

\begin{figure}
    \centering
    \includegraphics[width=\linewidth]{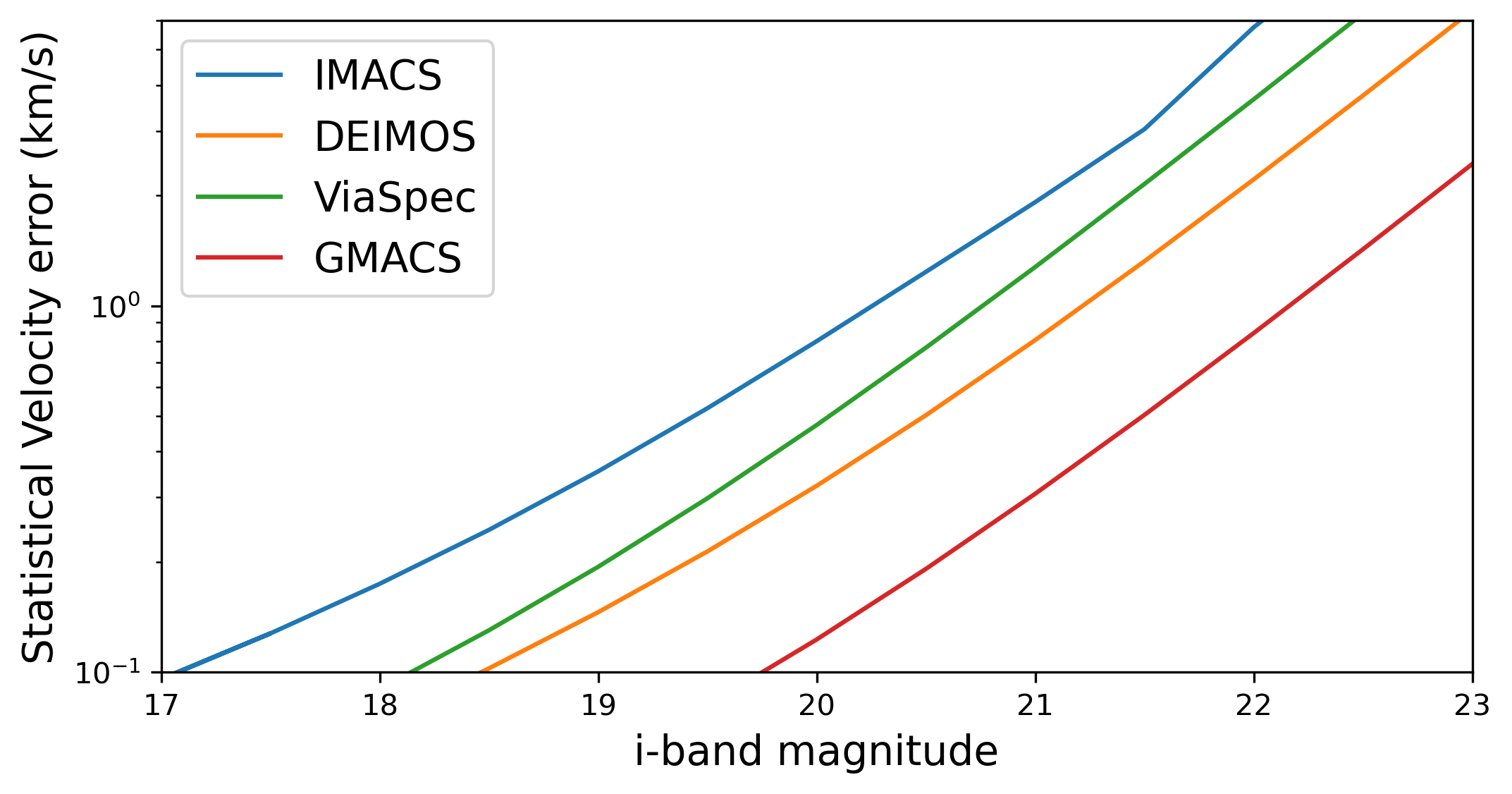}
    \caption{The magnitude-to-velocity uncertainty relations assumed through this work for Keck/DEIMOS, Magellan/IMACS, Magellan/Via, and GMACS for 6 hours of observation. These relations combine the effects of telescope mirror size, instrument resolution, and throughput. The kink in the IMACS curve comes from the extrapolation to fainter magnitudes.}
    \label{fig:verrsnr}
\end{figure}

In Figure \ref{fig:verrsnr}, we display the total assumed i-band-to-velocity uncertainty relations that we assume throughout this work for a 6-hour observation. This combines the magnitude-to-SNR and SNR-to-velocity uncertainty relations discussed above but does not include the systematic velocity uncertainty, which we specify during UFD generation.

\subsection{Observation Completeness Factor}
\label{subsec:completeness}
When comparing to actual observations, our simulations generate a factor of 1.3–5 more stars than are typically contained in literature catalogs for UFDs with similar properties and observing time. Possible sources for this discrepancy are:

\begin{enumerate}[nosep]
    \item Finite Slitmasks/Fields.\, Some of the UFDs have only been observed by a single slitmask, which will generally not include all of the stars in the system. This can occur due to crowding or if the system does not fit in a single pointing. Even for systems that have been observed with several masks, some stars may still not be included on the masks.
    \item Observational efficiency.\, Practical differences like weather, telescope observation patterns, etc.\ can decrease the SNR of the stars or affect their observability. Observers may also implement a more stringent magnitude limit than our SNR cut.
    \item Binaries and Variables.\, Membership for dwarf galaxies is typically determined after spectroscopy based on measured metallicity and radial velocities. Binaries with large radial velocity variations would be excluded from membership during this step.
    \item Milky Way Contamination.\, Milky Way foreground stars can be misidentified as members stars, thereby decreasing the efficiency for observing true member stars given the finite targeting constraints mentioned above.
\end{enumerate} These factors are difficult to comprehensively model, but each of them is expected to reduce the number of stars that are actually observed relative to the number that we naively generate (additional complexities arising from binaries are discussed in Section \ref{sec:disc}).

\begin{figure}
    \centering
    \includegraphics[width=\linewidth]{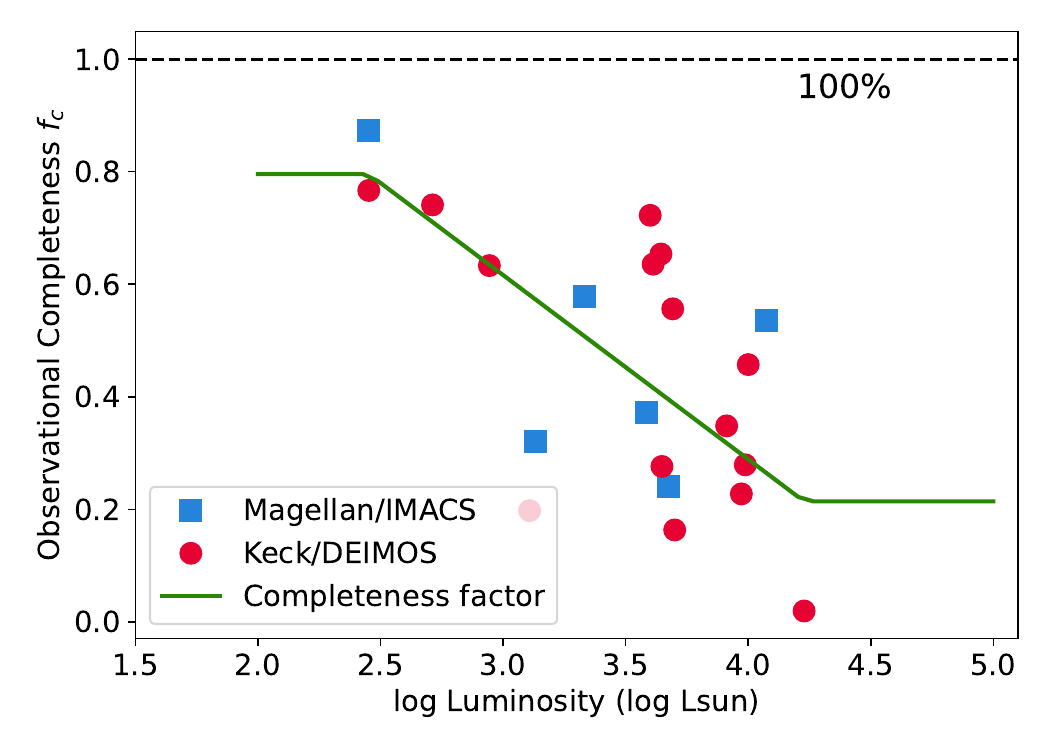}
    \caption{Observational completeness correction for spectroscopic observations. We estimated the observational completeness correction by comparing the number of stars predicted by our simulations to the number of stars actually observed. The solid lines represent the completeness factors we assume in our analysis.  }
    \label{fig:completeness}
\end{figure}

We account for this observational incompleteness empirically by defining a completeness factor, $f_c$, such that we select $f_cN$ stars from the $N$ stars generated by our simulation process (Section~\ref{subsec:ugali}). To derive this scaling factor, we select the IMACS-observed UFDs from the LVDB \citep{pace_local_2025} and the DEIMOS-observed systems from \citet{geha_keckdeimos_2026-1} and \citet{geha_keckdeimos_2026}. For these systems, we calculate the ratio between the number of member stars and the number that we naively generate for a system at the same luminosity, distance, and observing time. Because we exclude HB stars from our simulations, we also remove them from the number of member stars. Although the effective completeness factor should theoretically depend on the system’s distance and apparent magnitude, luminosity is the most obviously correlated observable. Thus, in Figure \ref{fig:completeness} we model it as a function of luminosity alone, fitting with linear regression and holding $f_c$ constant outside the fitted domain. Distance and apparent magnitude influence the expected number of visible stars, which $f_c$ then scales. 

\section{Projections}
In this section, we present projections for the ability to resolve UFDs with current, near future, and yet-undesigned instruments. We focus our study on Magellan/IMACS \citep{dressler_imacs_2011} and Keck/DEIMOS \citep{faber_deimos_2003} as representative examples of current instruments, since they cover nearly every UFD with spectroscopy in the literature. However, we note that other commonly used multi-object spectrographs include VLT/GIRAFFE \citep{pasquini_installation_2002}, Magellan/M2FS \citep{mateo_m2fs_2012}, MMT/Hectochelle \citep{szentgyorgyi_hectochelle_2011}, and AAT/2df+AAOmega \citep{saunders_aaomega_2004}. Our projections in Sec \ref{subsec:other_inst} can be used to broadly characterize the performance of other current instruments in resolving close and distant faint systems.
\label{sec:projections}

%, 4MOST \citep{de_jong_4most_2012}, and DESI \citep{desi_collaboration_desi_2016}
% are the spectrographs most often used for UFD follow-up and have the best combination of light-collecting power, throughput, and stability.

\subsection{Sample Size, Current Instruments}
\begin{figure*}
    \centering
    \includegraphics[width=\linewidth]{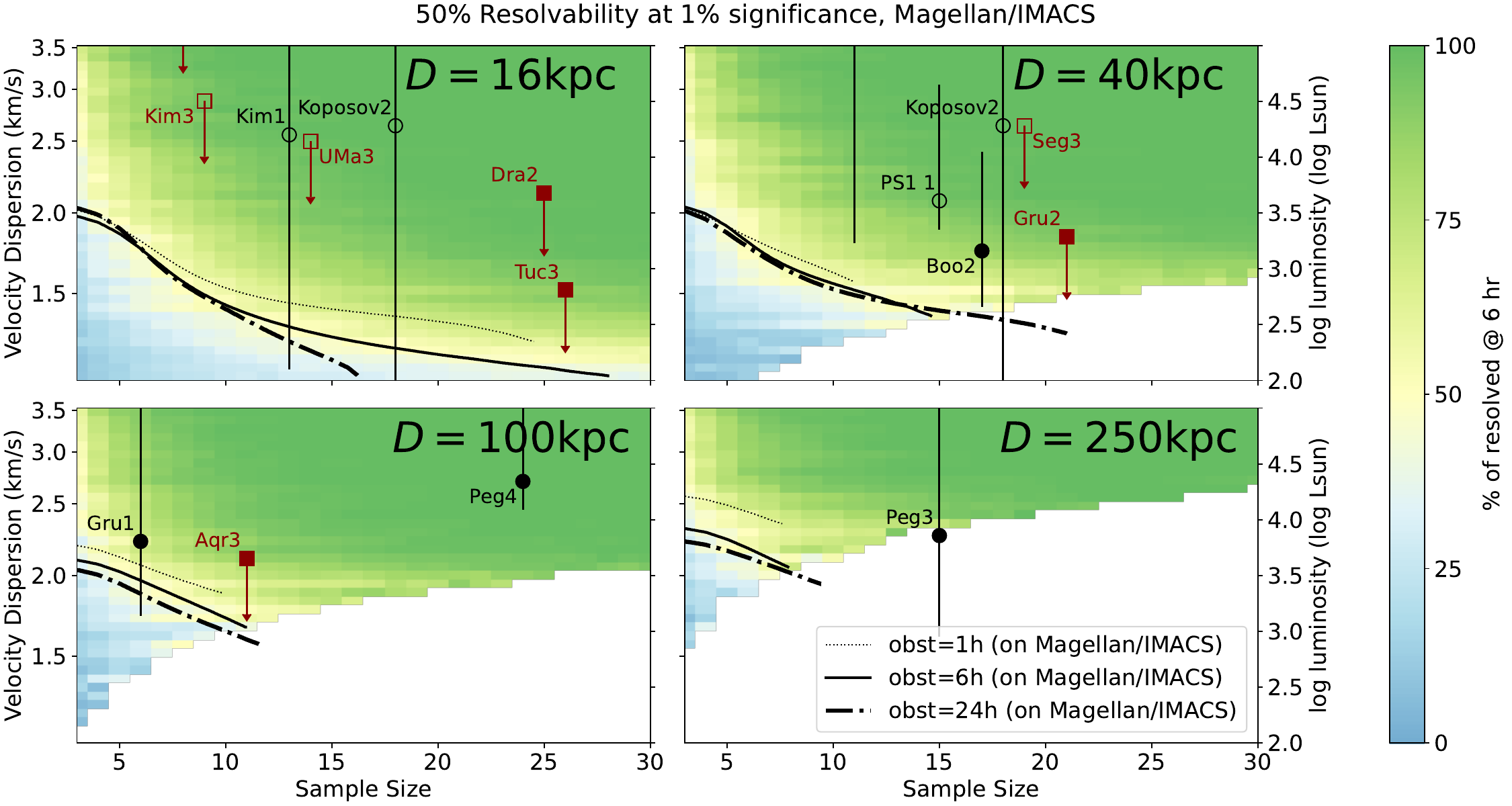}
    \caption{Resolvability projections for UFDs observed with IMACS at distances corresponding to distance moduli $\mu \in \{16, 18, 20, 22\}$. The lines trace where 50\% of our simulated UFDs were resolvable, using our hypothesis test with 1\% significance. The velocity dispersion here is derived from our luminosity scaling relation, hence we also present the vertical in terms of luminosity. IMACS- and DEIMOS-observed systems plotted here are positioned according to their measured velocity dispersions, not their luminosities: Solid markers are from our Table \ref{tab:ufds}, open are UFD candidates from \citet{cerny_chemodynamical_2026}, red squares are unresolved, black circles are resolved. The color contour corresponds to the resolvability that results from about a single night of observation. The white region corresponds to where our simulations could not generate enough stars for the given sample size, at the given luminosity. Our simulations link luminosity, halo mass, and half-light radius; however, individual unresolved systems may be more compact, reside in less dense halos, or be clusters, accounting for their positions on the plot.}
    \label{fig:ntable_deimos}
\end{figure*}

\begin{figure*}
    \centering
    \includegraphics[width=\linewidth]{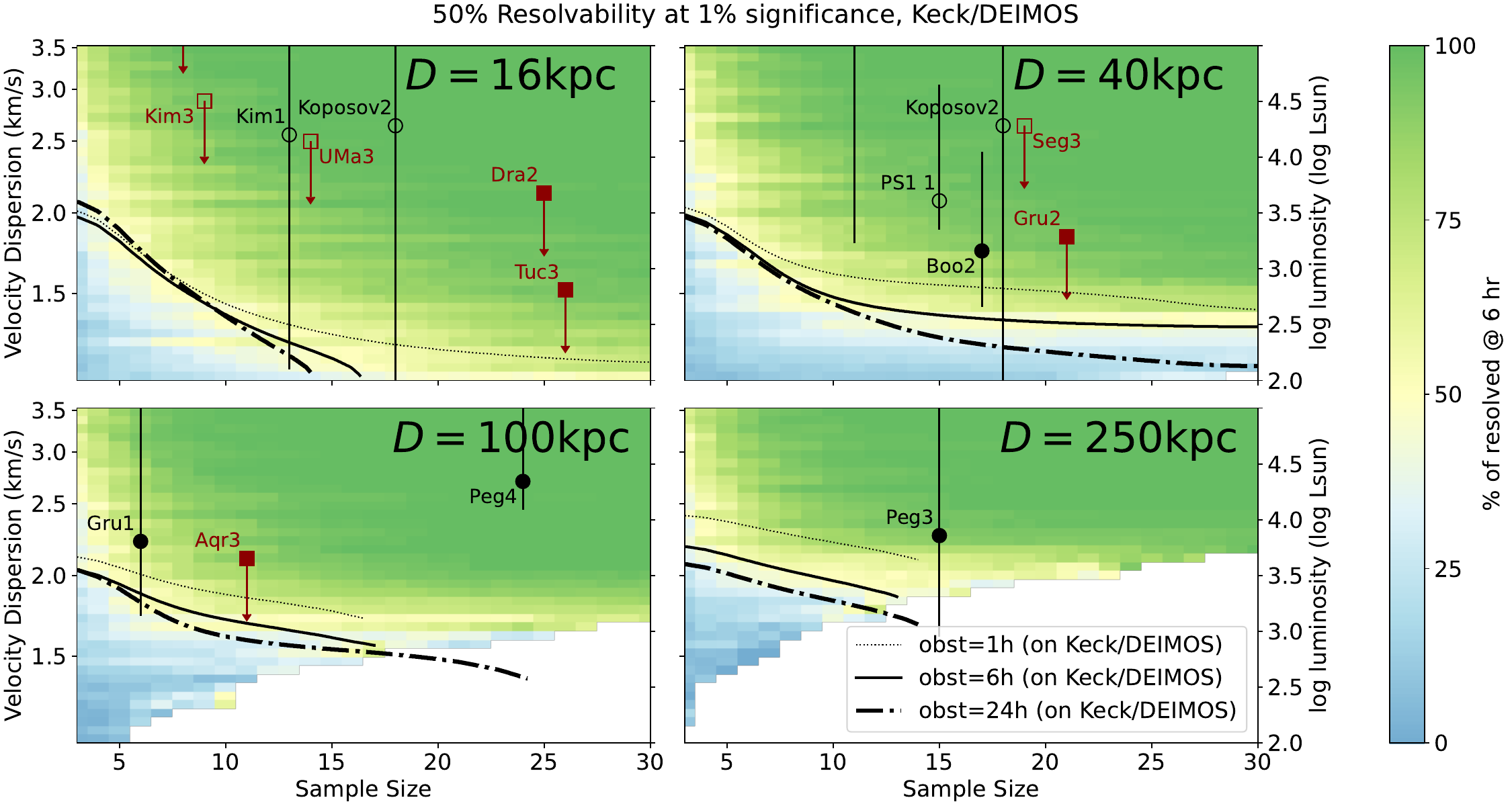}
    \caption{Same as Figure \ref{fig:ntable_deimos}, but for DEIMOS instead of IMACS. The difference in sensitivity primarily reflects the difference in the diameter of the Keck mirror ($10$\,m) compared to Magellan ($6.5$\,m).}
    \label{fig:ntable_imacs}
\end{figure*}

We first consider the resolvability of UFDs as a function of stellar sample size and velocity dispersion. To generate these projections, we produce ensembles of UFDs at combinations of distance modulus, $\mu$, and luminosity, $L_\mathrm{V}$: from each UFD, we sample the brightest $N$ stars and simulate observations of 1, 6, and 24 hours in ideal conditions. We then perform our hypothesis test to determine if such a sample is capable of resolving the system. Figures \ref{fig:ntable_deimos} and \ref{fig:ntable_imacs} display our projections for IMACS and DEIMOS respectively. The black lines correspond to the 50\% resolved contours---i.e., where half of the UFD samples at that combination of observational parameters are resolved. The colors display the percent of systems at the given observational parameters that can be resolved with 6 hours of observation, corresponding to about a single night: the green regions are where systems are resolved $>50\%$ of the time, and blue regions are where systems are resolved $<50\%$ of the time. The white excluded region denotes situations where the size of the stellar sample would need to be larger than could be observed for a dwarf with the given luminosity.

These figures can be used to estimate required exposure times to resolve a newly-discovered system. From photometric observations of a UFD candidate, one would determine the luminosity of that system. One can also estimate how many stars will be accessible in that system. With these estimates in hand, it is possible to locate the system in Figures \ref{fig:ntable_deimos} or \ref{fig:ntable_imacs} and determine where it resides. A UFD above the 1h line means that there is greater than a $50\%$ probability of resolving it with 1-hour of observation with the given instrument. A UFD above the 6-hour line has a greater than 50\% chance of being resolved with a full night of observing, and likewise with the 24-hour line. If the system lies below all of the lines, then there is a very poor chance of resolving it within 3--4 nights with this instrument. 

For UFDs with larger intrinsic dispersions and larger stellar samples, the velocity dispersion can be confidently resolved. However, resolvability decreases as the stellar sample size and velocity dispersion decrease. Both instruments have difficulty resolving UFDs with only 5 stars at an intrinsic dispersion of 2\,\kms. Velocity measurements in these cases are dominated by the systematic velocity uncertainty, so it is better to have multiple pointings for more complete membership of a candidate than to have a single, long exposure. For more distant systems, whose faint stars have large observational uncertainties, the depth of observations becomes the dominating factor.

Our projections show the observation time per star: if all stars can be observed simultaneously (i.e., on a single mask) then this requirement is trivially enforced, but if the stars are spread over multiple observations, then the total observation time would need to be increased by that factor. Our simulations only consider true members of the system, which can be difficult to determine before spectroscopy. Thus, the number of targets in real observations will likely need to be larger than our predictions to account for non-member contamination.

From our significance level, we expect 1\% of zero-dispersion star clusters to misclassified as resolved, regardless of where they sit in this parameter space. This could explain some of the unresolved systems (particularly, those from \citealp{cerny_chemodynamical_2026}) in Figures \ref{fig:ntable_deimos} and \ref{fig:ntable_imacs}. We also do not account for halo mass or half-light radius variability in our luminosity scaling relations; UFDs with smaller dispersions than we predict would be more difficult to resolve. Draco 2 \citep{longeard_pristine_2018}, Tucana 3 \citep{simon_nearest_2017}, and Grus 2 \citep{simon_birds_2020} are all suggested to have some degree of tidal stripping, which would decrease their velocity dispersions \citep{penarrubia_tidal_2008}. Aquarius 3 \citep{cerny_discovery_2025} lacks evidence of tidal stripping, but falls in a region on the plot where not all UFDs are resolved.

\subsection{LSST systems, Current Instruments}
\label{sec:discussion}
\begin{figure*}
    \centering
    \includegraphics[width=\linewidth]{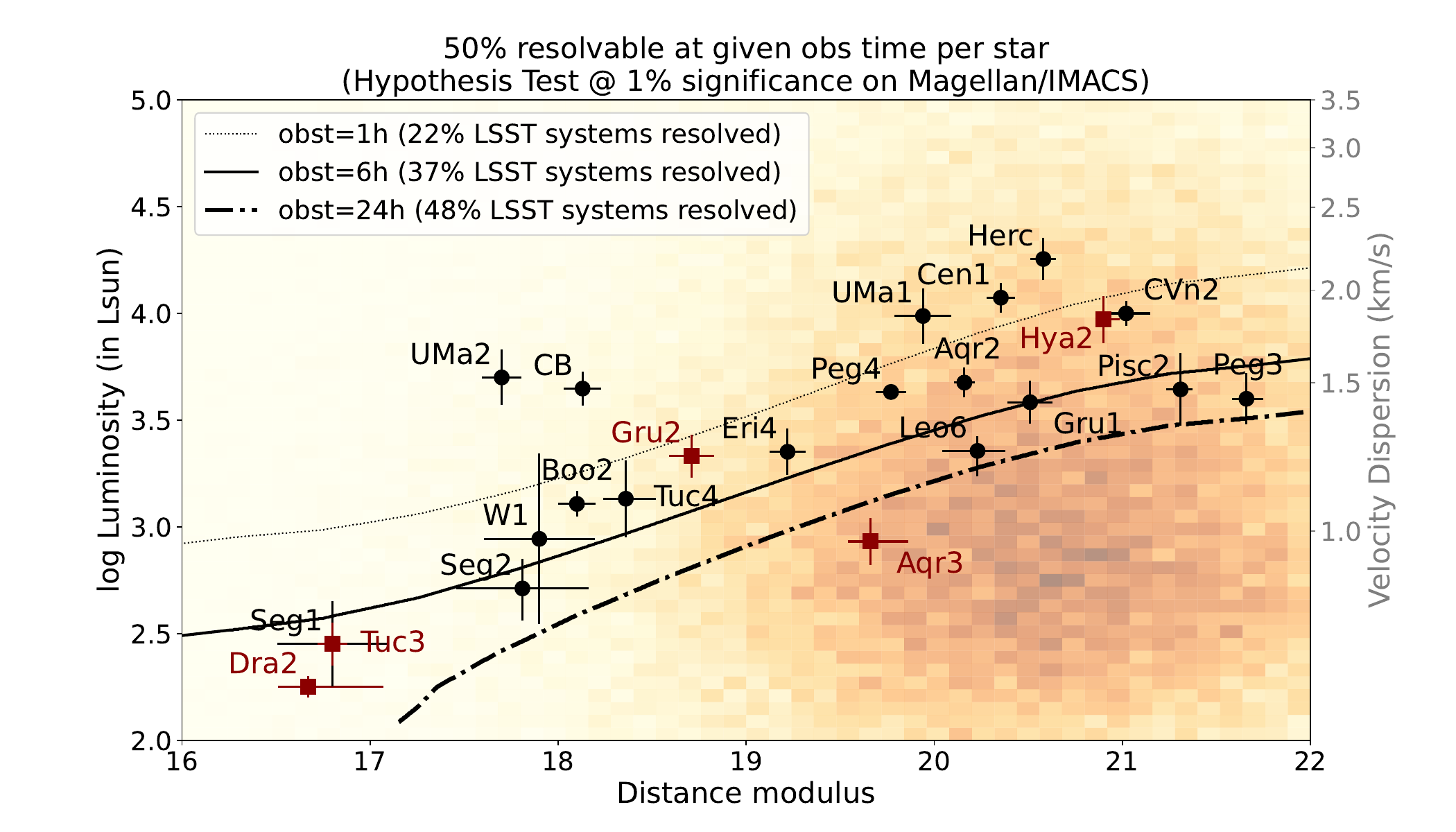}
    \caption{ The 50\% resolvability curves for IMACS, as a function of luminosity and distance modulus. IMACS- and DEIMOS-observed systems are plotted for comparison, in black (red) when (un)resolved according to our Table \ref{tab:ufds}. The orange histogram represents the satellite population that is predicted to be detected in LSST \citep{tsiane_predictions_2025}. The dispersions presented on the second vertical axis come from our luminosity scaling relation, so do not directly correspond to the dispersion of the plotted systems.}
        \label{fig:imacs projection}
\end{figure*}

\begin{figure*}
    \centering
    \includegraphics[width=\linewidth]{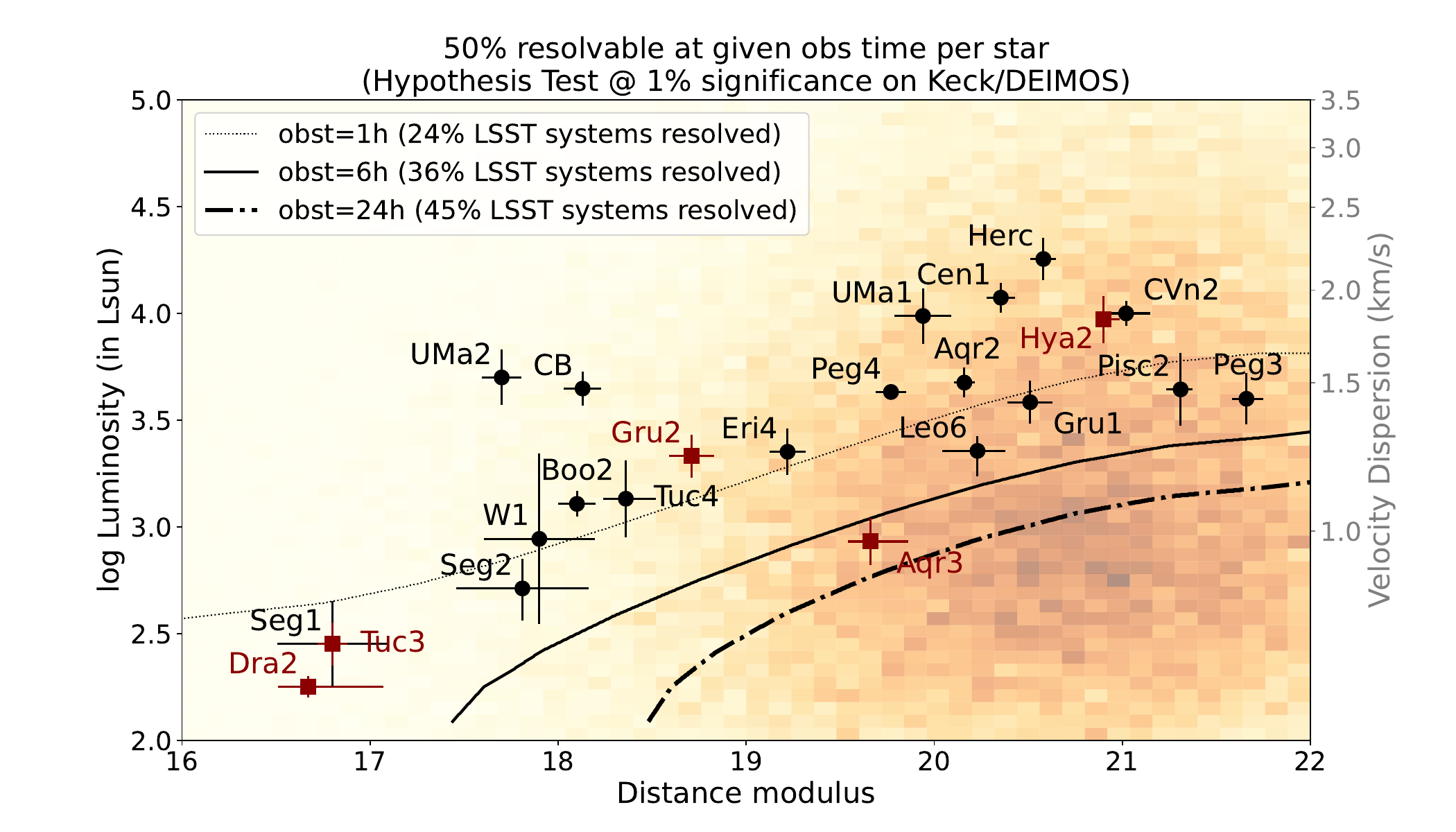}
    \caption{Same as Fig. \ref{fig:deimos projection} but for DEIMOS. The orange histogram represents the satellite population that is predicted to be detected in LSST \citep{tsiane_predictions_2025} with ${\rm Dec.} \sim -40^\circ$. Keck is larger than Magellan, so it can resolve more undiscovered LSST targets, despite reaching only a portion of the LSST footprint.}
    \label{fig:deimos projection}
\end{figure*}

%The NSF-DOE Vera C. Rubin Observatory's Legacy Survey of Space and Time (LSST) 
Rubin LSST is expected to discover many faint Milky Way satellites at large distances \citep[e.g.,][]{tollerud_hundreds_2008, hargis_too_2014, manwadkar_forward-modelling_2022}. We would like to estimate how many UFDs discovered by LSST will be resolvable with existing and future spectroscopic facilities. 
Simulating ensembles of UFDs varying distance and luminosity, we calculate the 50\% resolved lines for systems observed for 1, 6, and 24 hours with Magellan/IMACS and Keck/DEIMOS (Figures \ref{fig:imacs projection} and \ref{fig:deimos projection}, respectively). 

We compare our contours to predictions from \citet{tsiane_predictions_2025} for the Milky Way UFDs that LSST will observe. This LSST prediction comes from applying a LSST observational selection function to a UFD population predicted from a galaxy--halo model calibrated on data from DES and PS1 \citep{nadler_milky_2020}. \citet{tsiane_predictions_2025} apply an explicit cut at $M_V < 0$ (${\sim}100\, L_\odot$), where we truncate our plots. This distribution is displayed in right plot of in Figure 9 of \citet[in $M_V,D$ instead of $L_\mathrm{V},\mu$]{tsiane_predictions_2025} and has discrete structure because of the limited number of cosmological N-body simulations that were used to generate the Milky Way satellite population realizations. In our plots, we smooth this distribution to a model of the form
\begin{align*}
    \mathbb{P}(L_V,\mu) &= \mathbb{P}(\mu | L_V)\,\cdot\mathbb{P}(L_V) \\
    &\sim \mathcal{N}_s(\mu_1,\sigma_1,\alpha_1(L_V))\,\cdot \mathcal{N}_s(\mu_2,\sigma_2,\alpha_2)
\end{align*}
where $\mathcal{N}_s$ is a skew-normal function with skewness $\alpha$. LSST is expected to discover systems that are fainter and more distant than currently known satellites.

In Figure \ref{fig:imacs projection}, we show that 6 hours on IMACS will only be able to access 37\% of LSST-detectable systems, and 24 hours will be able to access 48\%. \citet{tsiane_predictions_2025} predict that $89\pm 20$ satellite galaxies will be observable with LSST, with 36 of those systems, or $40\pm 12\%$, already having been identified. This suggests that almost all of the systems in LSST that IMACS can resolve in a single night have already been identified.

While Keck has a larger primary mirror than Magellan, a large portion of the LSST footprint is below ${\rm Dec.} \sim -40^\circ$, which is inaccessible to Keck. This limits the ability of Keck to study the LMC and its associated satellites. In Figure \ref{fig:deimos projection}, we compare resolvability with DEIMOS to an LSST projection where the footprint is truncated below ${\rm Dec.} = -40^\circ$. Because of the restricted footprint, 36\% of the LSST-detectable satellites will be resolvable with 6 hours of observing, only increasing to 45\% when observed for 24 hours. This is comparable to IMACS even though DEIMOS is a more capable instrument. Only 21 satellite galaxies ($24\pm 6\%$ of LSST systems) have already been identified in the LSST footprint above ${\rm Dec.} = -40^\circ$, suggesting that DEIMOS will be an important instrument for the follow up of systems discovered by LSST. This does not take into account the fact that low-elevation observations would worsen velocity precision, and by extension, resolvability. Current facilities will be required to invest large amounts of time into individual systems in order to resolve them.

\begin{figure*}
    \includegraphics[width=\linewidth]{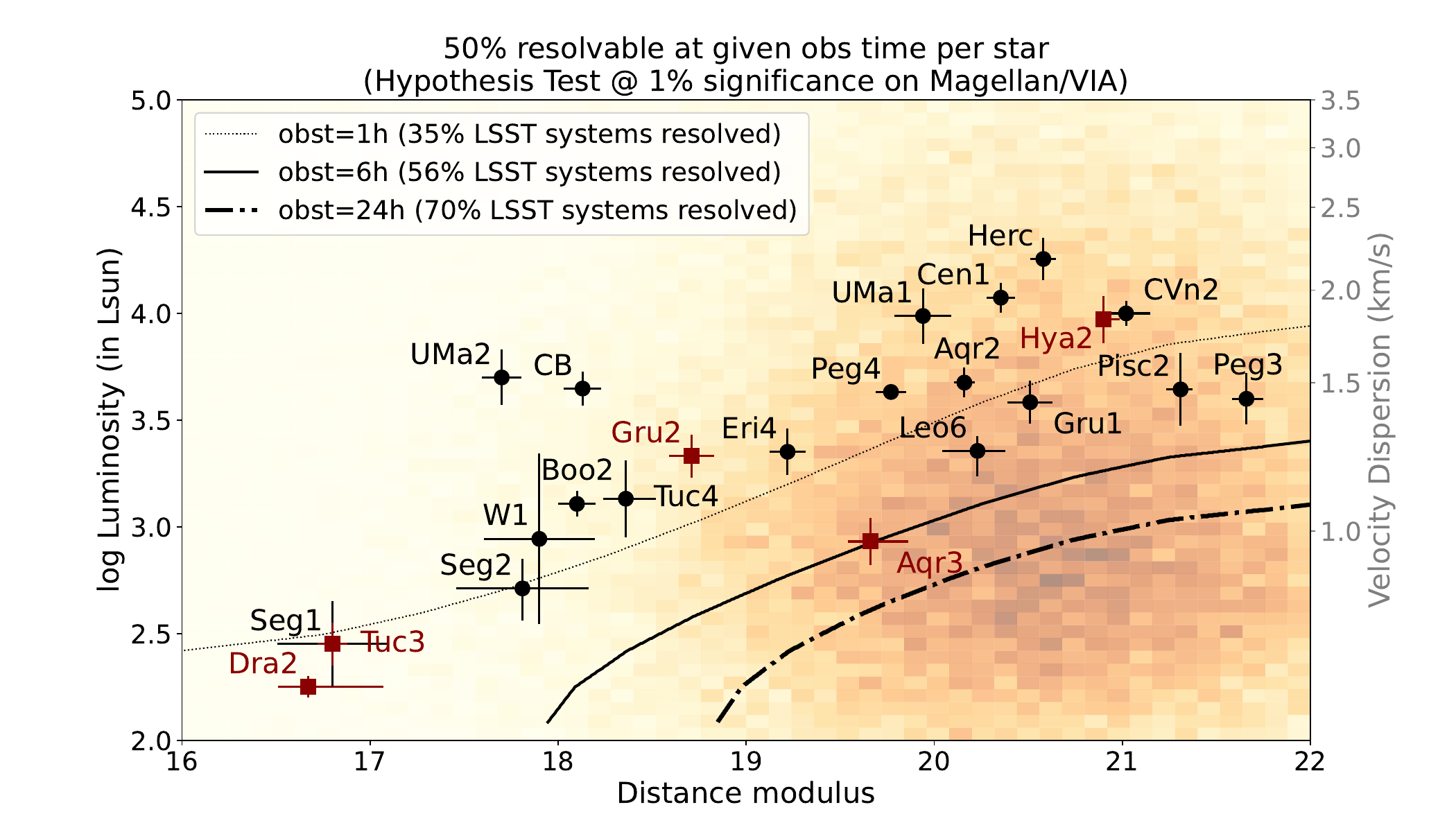}
    \caption{Similar to Figures \ref{fig:imacs projection} and \ref{fig:deimos projection}, we show the 50\% resolvability curves for Viaspec, which assumes a systematic velocity uncertainty of $0.1$\,\kms.}
    \label{fig:viaprojection}
\end{figure*}

\begin{figure*}
    \includegraphics[width=\linewidth]{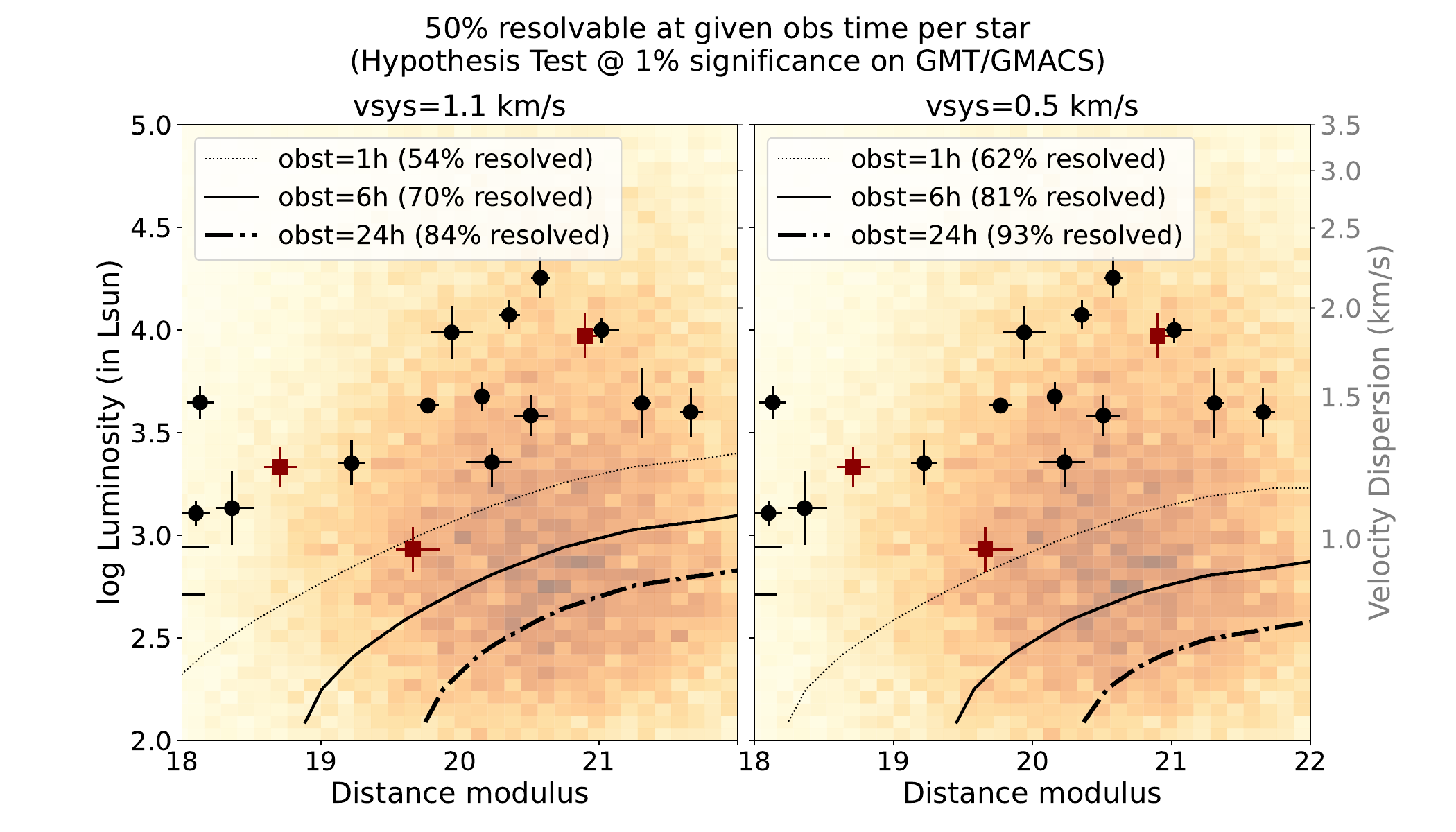}
    \caption{Similar to Figures \ref{fig:imacs projection}, \ref{fig:deimos projection}, and \ref{fig:viaprojection}, we show the 50\% resolvability curves for a GMACS with a 1.1\,\kms and GMACS+MANIFEST 0.5\,\kms systematic velocity uncertainty. Both assume performance similar to DEIMOS, except with 5 times larger mirror, but right has an additional 80\% throughput reduction due to MANIFEST fibers.}
    \label{fig:gmacsprojection}
\end{figure*}

\subsection{LSST systems, Near-Future Instruments}
While Magellan/IMACS and Keck/DEIMOS currently represent the state-of-the-art when it comes to measuring radial velocities of faint halo systems, over half of the LSST-detectable UFDs will be inaccessible to these instruments. However, the next generation of telescopes and instruments are already in advanced phases of design and construction. Viaspec and GMACS are two near-future spectrographs designed with UFD followup as a primary science goal. In a similar way to Section \ref{sec:discussion}, we project the abilities of Viaspec and GMACS to resolve LSST UFDs. 

Viaspec (the multi-object spectrograph for the Via project) will be a high-resolution ($R\approx 15{,}000$) optical fiber-fed spectrograph aiming for 0.1\,\kms velocity precision. It will achieve this through a combination of improved instrument stability, throughput, calibration, and modeling astrophysical sources of RV jitter (e.g., unresolved binaries, convective surface motion). Viaspec is scheduled to be deployed in 2027 on the 6.5m MMT (in Arizona) and Magellan/Clay (in Las Campanas, Chile) telescopes \citep{the_via_collaboration_via_2026}.

To model Viaspec radial velocity uncertainties, we assume a magnitude-to-SNR relation similar to VLT/FLAMES+GIRAFFE,\footnote{based on \url{https://www.eso.org/observing/etc/bin/gen/form?INS.NAME=GIRAFFE+INS.MODE=spectro}} except with 65\% of photons due to the smaller mirror (compared to 8.2\,m of VLT). We convert the SNR to a velocity uncertainty using the same relation from \citet{simon_kinematics_2007}, but assume 2.5 times smaller uncertainties due to a 2.5 times higher resolution than DEIMOS \citep{hatzes_doppler_2019}. The resulting magnitude-to-velocity uncertainty relation is plotted in Figure \ref{fig:verrsnr}. This relation sits between those derived for Viaspec from red giants with stellar metallicities $-2 <\mathrm{[Fe/H]}<-3$ (V. Chandra priv. comm.). Figure \ref{fig:viaprojection} shows the 50\% resolvability contours for UFDs observed with Viaspec. We predict that about 1.5 times as many systems will be resolvable in 6 hours with Via (56\% of LSST systems) than with IMACS (36\%). In particular, the high precision of the radial velocity uncertainties that Viaspec will enable it to resolve close, faint UFDs ($\mu<18$, $\sim40$\,kpc), rather than the $L\sim 10^2-10^3 \,L_\odot$ systems at the virial radius. However, our predictions are overly optimistic for these nearby systems because we do not model effects of main-sequence stars and fiber crowding. Our simulated uncertainties are a good approximation for galaxies with $M_\star>10^3\, M_\odot$, whose stellar velocities are dominated by red giant branch stars; smaller systems will require the observation of main-sequence stars with uncertainties around two times larger than we simulate \citep{the_via_collaboration_via_2026}. 

GMACS (the GMT Multi-object Astronomical and Cosmological Spectrograph) will be a first light instrument for the 25.4-m-diameter Giant Magellan Telescope at Las Campanas Observatory \citep{fabricant_gmacs_2025}. GMACS will operate with the MANIFEST fiber feed \citep{zafar_expanding_2026}. Though the resolution of GMACS alone will only be $R\sim 1000-6000$, MANIFEST is currently projected to have image slicing IFU units that will increase the spectral resolution to $R\sim 9000$ to 13000 around the calcium triplet \citet{zafar_enhancing_2024}.

% GMACS alone will provide R ~ 1000-6000 for a 0.7" slit. That is not really enough to provide the precision needed.
% I think that Alex Ji spoke to Febrcant and fond that the best way to achieve higher resolutions with GMACS is using MANIFEST to feed a set of gratings tuned to work with MANIFEST explicitly. The current plan includes a high resolution grating centered at  the Ca triplet, that provide R=10375 at that rest wavelength. The design in the blue side does not include a similar grating, but could if there is a scientific need.

Because GMACS is still being designed, we consider two scenarios in Figure \ref{fig:gmacsprojection}. Scenario 1, on the left, is GMACS only, with performance analogous to DEIMOS (except 5 times larger mirror) and a systematic velocity error of 1.1\,\kms. Scenario 2, on the right, is GMACS+MANIFEST with $R\sim 13\,000$ and a systematic velocity error of 0.5\,\kms. GMACS+MANIFEST is expected to have a better velocity uncertainty than DEIMOS due to its higher resolution (roughly twice that of DEIMOS) and its increased stability from being fiber-fed.  For MANIFEST, we assume a fiber throughput of 80\% \citep[D. Fabricant, priv. comm., also e.g.,][]{desi_collaboration_desi_2016,sharp_long-term_2013}. Figure \ref{fig:gmacsprojection} shows the 50\% resolvability contours for UFDs observed with GMACS; the left and right panels demonstrate the impact of the systematic velocity uncertainty on UFD resolvability.

At the same systematic velocity uncertainty as DEIMOS (Figure \ref{fig:gmacsprojection}; left panel), GMACS would be able to resolve more systems, including those at lower declinations, because the larger mirror would allow for deeper observations in the same amount of time. However, the major impediment in resolving systems is the systematic velocity floor of 1.1\,\kms. This is clear when considering a lower systematic velocity uncertainty of 0.5\,\kms (Figure \ref{fig:gmacsprojection}; right panel), which would allow GMACS to resolve 81\% of LSST UFDs with 6 hours each of observation, and 93\% with 24 hours each of observation. Therefore, reducing the systematic uncertainty of GMACS while retaining high throughput would be the most beneficial for UFD science.

\subsection{Other instruments}
\label{subsec:other_inst}

There are many forthcoming instruments on telescopes that span a range of sizes, including VLT/MOONS (\citealp{cirasuolo_moons_2020}, with an 8\,m primary), Spec-S5 (\citealp{besuner_spectroscopic_2025}, with two 6\,m primary mirrors), the multi-object spectrograph of the WST (\citealp{mainieri_wide-field_2024}, with a 12\,m primary), and ELT/MOSAIC (\citealp{pello_mosaic_2024}, with a 39\,m primary). As such, we generalize our results by considering instruments with a range of specifications. We focus on two characteristics: the light-collecting power and the systematic velocity uncertainty. In our model, we assume that the power to collect photons scales with the mirror area, using DEIMOS as a fiducial base.  Instrument throughput is another consideration that directly scales with light-collecting power, and it is generally the case that new instruments have higher throughput than older ones. For all UFDs in these projections, we use the DEIMOS magnitude-to-velocity error curve and scale the observation time as a proxy for mirror area (applying 80\% throughput reduction, assuming the instruments are fiber-fed). We directly vary the systematic velocity error during mock observation.

\begin{figure*}
\centering
    
    \includegraphics[width=\linewidth]{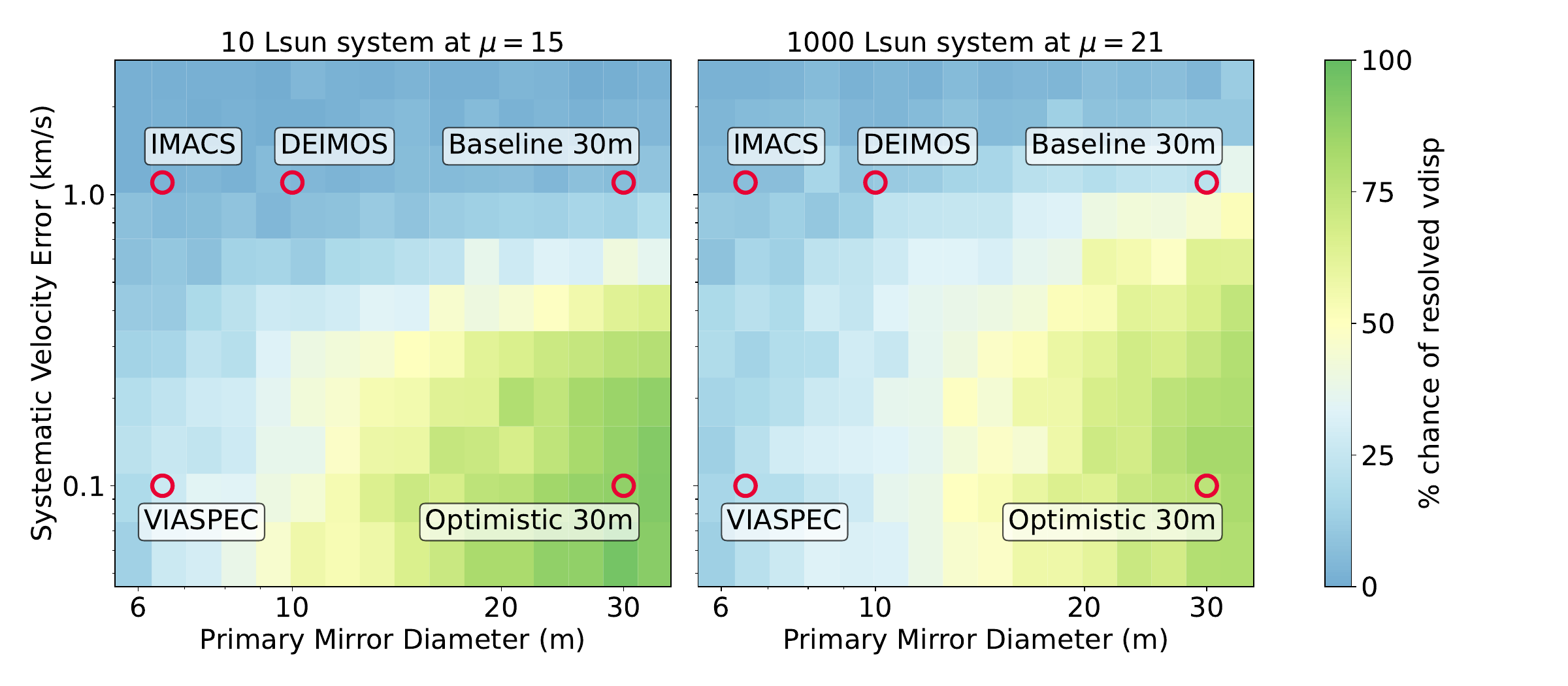}
    \caption{Projection for the resolvability of UFDs with current and future telescope/instrument configurations as a function of telescope mirror diameter and instrumental systematic velocity uncertainty. The left plot represents a system like Ursa Major III/Unions 1 \citep{smith_discovery_2024} with  $D\sim10$\,kpc and $L=10$\,$L_\odot$. The right plot represents the faintest systems ($L = 10^3\,L_\odot$) that may be resolved at a distance comparable to the Milky Way's virial radius ($D\sim250$\,kpc) with the most optimistic telescope and instrument configuration. Several current and near-future telescope/instruments are indicated.}
    \label{fig:futurefac}
\end{figure*}

In Figure \ref{fig:futurefac}, we present projections for two characteristic UFDs: one for a system like Ursa Major III/UNIONS I \citep{smith_discovery_2024} at a luminosity of $\sim 10\,L_\odot$ and a distance of $\mu=15$, representing the smallest system that has been detect at a distance of $D \sim 10$\,kpc, and another for a system $\sim 1000\,L_\odot$ at $\mu=21$, representing the least luminous system we will be able to detect near the virial radius of the Milky Way ($\mu\sim 22$; $D \sim 250$\,kpc). These projections assume 6 hours ($\sim$1 night) of observation. Each cell in this plot shows the percentage of UFDs that are resolvable with the given observational setup. The ability of these instruments to resolve the UFDs increase with larger mirror diameter and smaller velocity uncertainty, correspond to the abilities of these instruments to measure more precise velocities.

\begin{figure}
    \centering
    \includegraphics[width=\linewidth]{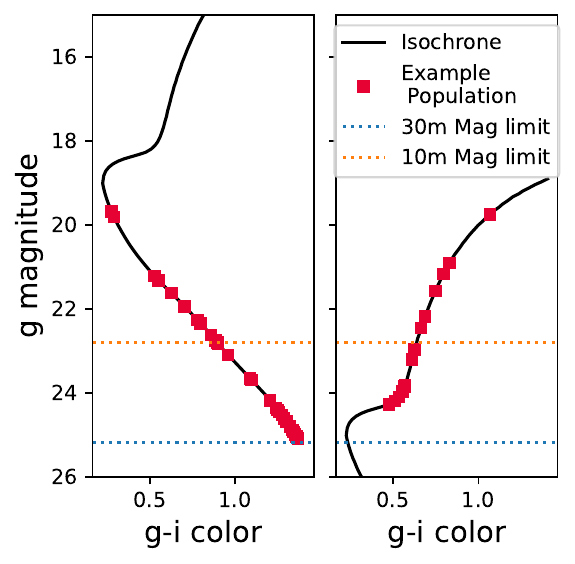}
    \caption{Example idealized color--magnitude diagrams for the systems studied in Figure \ref{fig:futurefac}. The isochrones come from \citet{bressan_parsec_2012}, populated with a Chabrier IMF \citep{chabrier_galactic_2003}. We plot magnitude limits from our simulations for comparison.}
    \label{fig:examplecmd}
\end{figure}

We show example color magnitude diagrams for these two systems in Figure \ref{fig:examplecmd}. The observational sample of the nearby system (left plot) is dominated by main-sequence stars, many of which are inaccessible to current instruments (falling below the 10m telescope magnitude limit). This is consistent with systems in \citet{cerny_chemodynamical_2026} (Ursa Major III/UNIONS 1, Kim 3). We note that our simulated uncertainties are optimistic in this regime, since they are based on observations of red giant stars, not main-sequence stars. The more distant system sample (right plot) is composed of red giants, while the majority of the main-sequence stars are too faint to be detected by our simulated observations. 

Magellan/IMACS and Keck/DEIMOS struggle to resolve the close UFDs that we examine. This is consistent with the fact that the velocity dispersion originally claimed for Ursa Major III/Unions 1 was anomalously high given its stellar mass and our current understanding of the galaxy--halo connection. It was subsequently found that the dispersion measurement was being inflated by a binary star \citep{cerny_no_2025}. Furthermore, there is little chance of resolving the more distant system given current facilities.

In contrast, the lower systematic floor expected for Viaspec improves the chances of resolving Ursa Major~III/UNIONS 1-like systems, though the probability is still $<50\%$. Viaspec is also unable to resolve the distant system, since the sizes of the Magellan and MMT mirrors limit the number of stars that are observable. 

Interestingly, our projections suggest that increasing mirror diameter alone is not enough to confidently resolve either of the extreme systems that we consider. In both cases, a 30-meter-class telescope hosting an instrument with a systematic velocity uncertainty of 1.1\,\kms (comparable to DEIMOS and IMACS) will fail to resolve both faint and distant systems, though this ``Baseline 30m'' design is able to resolve a $1000\,M_\odot$ system at $\mu=21$ ($D\sim 160\,$kpc) $>25\%$ of the time. In contrast, if the systematic uncertainty can be reduced to 0.1\,\kms, then such a configuration is capable of confidently resolving systems like Ursa Major~III/Unions 1 and can resolve $>50\%$ of faint systems near the Milky Way virial radius. 

Intermediate-scale telescopes would be more capable at resolving nearby ultra-faint compact systems than extremely distant $1000 L_\odot$ systems. As an example, we point to  the forthcoming 12\,m Large Optical Telescope that will be built in western China \citep{cui_introduction_2018,pei_study_2025}. While a northern hemisphere observatory would not be as useful for following LSST-detected UFD candidates, it would be capable of observing northern-hemisphere systems discovered by UNIONS and Euclid, especially if its spectrographs have a smaller systematic velocity uncertainty ($<0.3$\,\kms). Such an instrument would also be more capable than Viaspec at observing smaller systems near the virial radius, though not as capable as a 30-meter facility with the same systematic velocity uncertainty. The 12-m Wide-field Spectroscopic Telescope (WST, \citealp{mainieri_wide-field_2024}) planned for the southern hemisphere may be useful in following LSST-detected UFD candidates; however, its wide field-of-view will make it difficult to observe crowded compact systems. They plan for an IFU to be the main instrument for dwarf followup. 

\section{Discussion}
\label{sec:disc}

\subsection{Binaries}
\label{sec:binaries}
For simplicity, our simulations exclude stellar binaries; the rate of binaries varies drastically between UFDs \citep[e.g.,][]{minor_binary_2013,spencer_multi-epoch_2017}, though it has been suggested that either binary rate or binary radial velocities decreases with galaxy mass \citep{spencer_binary_2018}. However, the smaller the object and its dispersion, the larger the impact binaries can have \citep{mcconnachie_revisiting_2010,gration_stellar_2025,arroyo-polonio_estimating_2026}. There is also a concern that low metallicity systems would have higher close binary fractions \citep{moe_close_2019}. RGB binaries in UFDs have been previously measured with typical periods of months and orbital velocities of $\sim30$\,\kms \citep{koposov_accurate_2011,koch_spectroscopic_2014,ji_high-resolution_2016}.

Inflated velocity dispersion measurement can lead to false positives for resolved systems \citep[e.g.,][]{cerny_no_2025}. Our simulations do not include binaries, meaning that all of the stars in each synthetic UFD trace its gravitational potential. To connect projections with actual observation, one might factor in binaries by assuming a binary fraction and increase the number of stars one aims to observe to compensate for it. Then, one could discard velocity outliers and still have enough stars to resolve a dispersion.

Ideally, one would plan to revisit stars in order to look for velocity variability before incorporating those stars into dispersion analysis. With multiple epochs of observation, one could characterize the binaries and incorporate the systemic radial velocity into the velocity dispersion calculation, \citep[e.g., the JoKer,][]{price-whelan_joker_2017,buttry_stellar_2022, minor_robust_2019} however, multi-epoch observation greatly increases the required telescope allocation  and it can be difficult to fit binary periods. In future work, we hope to incorporate binaries into our model.

\subsection{Metallicity Dispersions}

We have phrased our technique in the language of velocity dispersions, but in principle, a similar argument could be made for resolving {\it metallicity} dispersions. Typical UFD metallicity dispersion range between $\sigma_\mathrm{[Fe/H]}\approx 0.3-0.7$\, dex \citep{simon_kinematics_2007, simon_faintest_2019,fu_metallicity_2023}. A spread in metallicities signifies that a system has a gravitational potential that is deep enough to retain enrichment from multiple generations of star formation, implying the presence of a dark matter halo in ultra-faint systems \citep{willman_galaxy_2012}. 

Binaries are not as problematic for measuring metallicities, removing the need for multi-epoch data. Useful information can also be derived from samples of one or two stars (as was done in \citealp{cerny_chemodynamical_2026}), since UFDs are universally metal-poor \citep{kirby_uncovering_2008,kirby_universal_2013,salvadori_ultra_2009}.

A complication to this is that the metallicity distribution functions of galaxies are not expected to be Gaussian, so Equation \ref{eqn:thresh} cannot be used. One would instead have to derive an appropriate LHR threshold from simulations. Stellar velocity measurements are also generally more precise, more scalable with multi-object spectrographs, and can be determined for fainter stars than metallicity measurements.  

For the identification of individual systems, one would like to take an approach similar to that of \citet{cerny_chemodynamical_2026}: using multiple indicators (e.g., mean metallicities, metallicity dispersions, ages, mass-to-light-ratios, mass segregation, and chemical abundance patterns) to characterize candidates as UFDs or clusters. Joint hypothesis tests that combine metallicity and velocity dispersion \citep[e.g.,][]{walker_magellanm2fs_2015,koposov_snake_2018} could make UFD identification even more robust and may be necessary to conclusively identify all of the systems that we will find in near future surveys. 

\subsection{Purity and Completeness of UFD census}
Ultimately, systematizing the measurement of velocity dispersions is in service of making rigorous and robust statistical inferences about the dwarf galaxy population. One consequence of our statistical formalism is that we should expect that applying the same dispersion analysis to true star clusters would lead to 1--3\% of them being identified as dwarf galaxies. Thus, population-level analyses of UFDs should consider that there could be contamination in the measured UFD sample. This is relevant to analyses that are sensitive to the properties of the ``smallest dwarf galaxy'' \citep[e.g.,][]{dalal_excluding_2022}.
 
 The other, more interesting consequence, is that we expect there to be a selection function due to the resolvability of UFD dispersions. The fiducial Milky Way model from \citet{nadler_milky_2020} predicts $\sim100$ UFDs between the luminosity range of 100 and 1000\,$L_\odot$. From \citet{tsiane_predictions_2025}, we expect LSST to be able to detect $\sim28$ systems at these luminosities, $\sim 20$ at declinations accessible to Keck. As an estimate of completeness due to resolving velocity dispersions, we take the LSST UFD distribution from Figure \ref{fig:deimos projection} and convolve it with the resolvability at 6 hours; we find that $\sim27\%$ of systems detectable with LSST between 100 and 1000\,$L_\odot$ can be resolved with DEIMOS---this means we could expect to confirm $\sim5$ UFDs with DEIMOS with $M_v>-2.7$ in the region of LSST that DEIMOS can access. Repeating this for IMACS with Figure \ref{fig:imacs projection}, we get $\sim5$ UFDs.  While approximate, this makes it clear that current instruments are not be able to recover a large population of UFDs below $10^3\,L_\odot$, contributing to the paucity of confirmed UFDs at these low luminosities. However, there are observed UFDs in this regime, and they count towards these totals. Lastly, if we repeat this estimate for our GMACS projections (in Figure \ref{fig:gmacsprojection}) with a systematic velocity uncertainty of 0.5\,\kms, we get $\sim 15$ UFDs resolvable within the LSST footprint.

 We note that \citet{manwadkar_forward-modelling_2022} predict 440 total MW satellites, twice as many as \citet{nadler_milky_2020}, with around 300 of those galaxies falling between 100 and 1000\,$L_\odot$.  This is because the GRUMPY dwarf galaxy formation model generates more dwarfs at lower halo masses. With this model, we would expect to confirm more galaxies from LSST, though the overall completeness of the UFD sample would be lower because those galaxies would have smaller dispersions. The incompleteness of the faint end is relevant to the derivation of dark matter constraints from population-level measurements of velocity dispersions, which are particularly sensitive to the behavior of the smallest UFDs \citep[e.g.,][]{sullivan_minimum_2018,esteban_milky_2024}. 

\section{Conclusions}
\label{sec:conclusion}

 Ultra-faint dwarf galaxies probe the physics of galaxy formation and evolution, and may illuminate the physics of dark matter. These pursuits require us to accurately identify increasingly faint UFDs based on their velocity dispersions with limited observational resources. We have projected the ability to measure stellar velocity dispersions in UFDs using current and future spectroscopic facilities. To enable this work, we formalized a likelihood-ratio-based hypothesis test as a standardized technique to distinguish resolved and unresolved systems. 
 
 The likelihood ratio test statistic is directly calculated from the data and compared to a threshold value to determine whether the data significantly supports the hypothesis of a nonzero velocity dispersion. We apply this hypothesis test to several known dwarf galaxies and find that the wide variety of criteria used within the community to claim resolved velocity dispersions generally agree with the likelihood ratio hypothesis test with a threshold set at 1--3\% significance level. 

Using kinematic simulations, we project velocity dispersion resolvability as a function of member star sample size, intrinsic velocity dispersion, and distance for Keck/DEIMOS and Magellan/IMACS, the two instruments that are most commonly used for UFD followup. These projections can help inform the design of spectroscopic followup observations of UFD candidates.

 We also project that multi-night Keck/DEIMOS observations will be able to resolve some of the UFDs that are expected to be discovered by LSST, despite being unable to access the full footprint. IMACS will be less useful because it has already found most of the systems it is capable of resolving. Viaspec and GMACS, two near-future spectrographs designed with UFD characterization as a primary science goal, will be necessary to discover many of the new systems from LSST. In particular, for GMACS to have the largest impact in resolving velocity dispersions, it needs to deliver the smallest systematic velocity uncertainties possible. Future instruments with 1\,\kms velocity precision mounted on 30-meter-class telescopes (e.g., GMT, TMT, E-ELT) will only provide a slight advantage compared to near-future high-precision instruments like Viaspec ($\sim$0.1\,\kms velocity uncertainty), while even a moderate-precision instrument ($\sim 0.3\,$\kms velocity uncertainty) on a 30\,m telescope could enable the least massive known systems ($L \sim 10\,L_\odot$) to be resolved. However, even in this optimistic case, the faintest and most distant Milky Way systems will remain challenging to resolve. 

~~

\section*{Acknowledgments}

We would like to thank Will Cerny for helpful discussions. We would like to thank Vedant Chandra and Daniel Fabricant for communications regarding Via and GMACS.

This work was partially supported by the National Science Foundation Astronomy \& Astrophysics Grants program award No.\ AST-2307126 and AST-2407526. This research was partially funded by a generous gift from Charles Simonyi to the NSF Division of Astronomical Sciences.

% Alex's Fermilab acknowledgment 
Fermilab is managed by FermiForward Discovery Group, LLC under Contract No.\ 89243024CSC000002 with the U.S.\ Department of Energy, Office of Science, Office of High Energy Physics. The United States Government retains and the publisher, by accepting the article for publication, acknowledges that the United States Government retains a non-exclusive, paid-up, irrevocable, world-wide license to publish or reproduce the published form of this manuscript, or allow others to do so, for United States Government purposes.

\appendix
\section{A. Derivation of Likelihood Ratio Null Distribution}
\label{app:deri}

In our hypothesis test, we calculate a likelihood ratio (LHR) test statistic that encapsulates how well the data supports an alternative hypothesis (nonzero dispersion) over the null hypothesis (zero dispersion). To interpret the LHR test statistic, $\lambda_\mathrm{lh}$, we compare it to the distribution of LHRs under the null case. If the test statistic exceeds $(1-\alpha)$\% of the null distribution, we say that the sample is resolved with $\alpha$\% significance. This is equivalent to saying that the same LHR threshold would false identify only $\alpha$\% of null cases. 

Under the conditions of applicability, Wilks' Theorem provides an analytic formulation for the distribution of $\lambda_\mathrm{lh}$ \citep{wilks_large-sample_1938}. In the case where the true parameter lies on the boundary of the parameter space, like $\sigma_v>0$, this distribution is described by Chernoff's extension \citep{chernoff_distribution_1954}. However, these analytical formulae are only guaranteed to apply in the case of large sample size, which is not true for UFD observations. This motivates us to derive the null distribution of the LHRs without making assumptions about the size of the sample. We will demonstrate that under the null hypothesis ($\sigma_v=0$), the LHR distribution is $\lambda_\mathrm{lh}=X_{N}-N-N\log(X_N/N)$ where $X_N\sim \chi_{N-1}^2$ and therefore Equation \ref{eqn:thresh} holds. 

In order to derive the distribution of the LHR test statistic, we make two simplifying assumptions: (1) that the velocities of the stars are sampled from a Gaussian distribution, $v_i\sim \mathcal{N}(\mu_v,e_i^2+\sigma_v^2)$, and (2) that the velocity uncertainties are the same for all stars, $e_i=E$. Using \texttt{StarSampler}, we find that first assumption is a good approximation for our choice of NFW matter density and Plummer stellar density \citep{navarro_universal_1997,plummer_problem_1911}. The latter assumption is not true when considering stars spanning a wide range of SNR. However, this assumption makes the calculation tractable, and we demonstrate with simulations that the resulting distribution adequately holds over the range of observational uncertainties in our study.

First, we rewrite Equation \ref{eqn:lh} with homoskedastic uncertainties as:
\begin{align}
2 \ln \mathcal{L} = -N\ln(2\pi)-N\ln (\sigma_v^2+E^2)-\sum_{i=1}^{N}\frac{(v_i-\mu_v)^2}{\sigma_v^2+E^2}
\end{align}
The maximum likelihood estimator, $\sigma_b=\sigma_v^\mathrm{MLE}$ maximizes this quantity (by extension, maximizing $\mathcal{L}$) and is given by:
\begin{align}
    \frac{\sigma_b^2}{E^2}+1&= \frac{1}{N}\sum_{i=1}^{N} \frac{(v_i-\mu_v)^2}{E^2}.
\end{align}
The right-hand side corresponds to the velocity dispersion we measure from our sample. The left-hand side can be decomposed into an uncertainty component (the second term) and a component attributed to the intrinsic velocity dispersion (the first term). This comes naturally from assuming our data and uncertainties are Gaussian. If the first term is smaller than the second term, $\sigma_b=0$ maximizes $\mathcal{L}$. Then we can rewrite Equation \ref{eqn:LHR}:
\begin{align}
    \lambda_\mathrm{lh}&=2\ln\mathcal{L}(\sigma_b)-2\ln\mathcal{L}(\sigma_v=0)\\
    &= -N\ln(\sigma_b^2+E^2)-\frac{E^2}{\sigma_b^2+E^2}\sum_{i=1}^{N} \frac{(v_i-\mu_v)^2}{E^2} +N\ln(E^2)+\sum_{i=1}^{N} \frac{(v_i-\mu_v)^2}{E^2}\\
    &=-N \ln\left(\frac{1}{N} \sum_{i=1}^{N} \frac{(v_i-\mu_v)^2}{E^2}\right) -N +\sum_{i=1}^{N} \frac{(v_i-\mu_v)^2}{E^2}\\
    &= X_N - N - N \ln(X_N/N),
\end{align}
where 
\begin{equation}
    X_N=\sum_{i=1}^{N} (v_i-\mu_v)^2/E^2.
\end{equation}

Lastly, for a null (zero-dispersion) system, $v_i\sim \mathcal{N}(\mu_v,E^2)$, so $(v_i-\mu_v)/E\sim \mathcal{N}(0,1)$. This makes $X_N$ the sum of squared normal distributions, or a chi-squared distribution. When evaluating real data, we derive $\mu_v$ from $v_i$, removing a degree of freedom from the chi-squared distribution, making $X_N\sim \chi_{N-1}^2$.

\begin{figure*}[h]
    \centering
    \includegraphics[width=\linewidth]{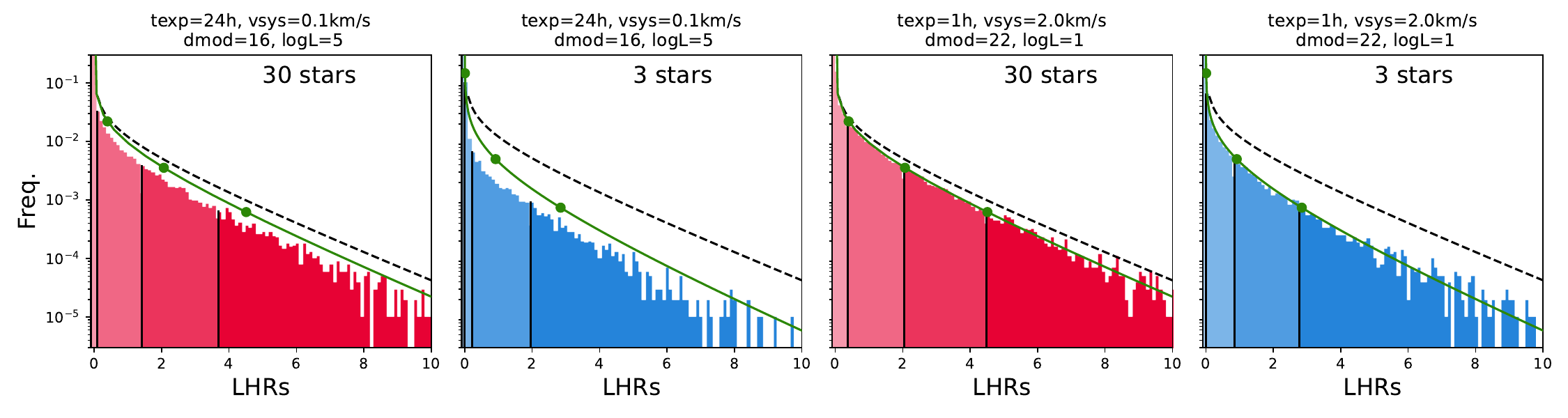}
    \caption{$\lambda_\mathrm{lh}$ distribution for the most extreme observational conditions we consider in this paper. The left two panels consider 30 star and 3 star samples from a nearby UFD ($\mu = 16$) observed for 24\,hours with a small systematic velocity uncertainty (0.1\,\kms). The right two panels consider a distant UFD with low stellar mass ($L = 10\,L_\odot$), observed for 1\,hour, with a large systematic velocity uncertainty (2.0\,\kms). The dots correspond to the 80th, 95th, and 99th percentile, which also qualitatively agree with the distributions (denoted by change in shade and vertical line). The black dashed lines correspond to the distribution predicted by Chernoff's extension to Wilk's Theorem \citep{chernoff_distribution_1954}.}
    \label{fig:wilks}
\end{figure*}

This derivation is approximate, particularly because of our assumption of homoskedasticity, but we use simulations to demonstrate that it is adequate for our analysis. The left two panels of Figure \ref{fig:wilks} represent most precise velocities we expect to observe in our analysis: $L=10^5\,L_\odot$ systems at a distance of $\mu=16$ (16\,kpc), with a systematic uncertainty of 0.1\,\kms, with 24 hours of exposure time. The right two panels represent the least precise uncertainties we expect, $L=10\,L_\odot$ systems at a distance of $\mu=22$ (250\,kpc), observed for 1 hour with a systematic uncertainties of 2\,\kms. 

We can see both that the LHRs of these cases are discrepant from the Chernoff's extension to Wilks' theorem (dashed line) and better align with the analytic distribution from Equation \ref{eqn:thresh} (solid green line). Moreover, the dots representing the 20\%, 5\% and 1\% significance levels match the 80th, 95th, and 99th percentiles that are empirically determined from the $\lambda_\mathrm{lh}$ distributions, as expected. The assumption of homoskedastic uncertainties is better in the cases with larger systematic velocity uncertainties, since the variation in the total velocity uncertainties is much narrower. We can see, however, that in the cases with smaller uncertainties, the percentiles of the null distribution are smaller than those from the analytic solution. This means that in these cases, Equation \ref{eqn:thresh} generates a slightly stricter threshold than is reflected in simulations. This means that our resolvability projections for high-precision instruments are slightly conservative.

\section{B. UFD systems}
\label{app:ufd}
Here we list the UFD systems that are considered in our analysis. The properties of the systems are sourced from the Local Volume Database \citep{pace_local_2025}, and the spectroscopic data come from a combination of \citet{pace_scaling_2019} (who compiled the data from their respective sources) and the DEIMOS archive \citep{geha_keckdeimos_2026-1,geha_keckdeimos_2026}. We prioritized the DEIMOS archive data for systems besides the classical dwarfs. For systems that exceed the hypothesis test threshold at a significance of 1\%, we report $\sigma_v$ derived using a Monte Carlo Markov Chain (MCMC) with a uniform prior. The reported values are the maximum {\it a posteriori} value and the uncertainties come from the maximum-density interval containing 68\% of the posterior distribution. For systems that fail to exceed the hypothesis test threshold, an upper limit is reported that contains 90\% of the posterior. The computation of these LHRs and velocity dispersions are included in a Jupyter Notebook in the \code{dsphsim} repository linked in Section \ref{sec:methods}.
\begin{table*}[h]
    \centering
    \caption{The full table of ultra-faint dwarf galaxies (UFDs) in our analysis}

\begin{tabular}[t]{c|c|c|c||c|c|c|c}
          UFD & Instrument &Source & Nstar &  $\alpha=1\%$& LHR & Decision & Dispersion\\ 
          
            & & & &$\Lambda_\mathrm{th}(N_\mathrm{star})$& $\lambda_\mathrm{LH}$ & & $\sigma_v$ \\\hline

Ant2 & AAT/AAOmega+2dF & \citet{ji_kinematics_2021} & 284 & 5.10 & 2916.47 & Resolved & $6.39_{0.34}^{0.28}$ \\
Aqr2 & Keck/DEIMOS & \citet{geha_keckdeimos_2026-1} & 8 & 3.78 & 26.69 & Resolved & $4.29_{1.20}^{2.50}$ \\
Aqr3 & Keck/DEIMOS & \citet{geha_keckdeimos_2026-1} & 11 & 4.00 & 0.00 & Unresolved & $<3.22$ \\
Boo1 & Keck/DEIMOS & \citet{geha_keckdeimos_2026-1} & 92 & 4.88 & 118.57 & Resolved & $3.30_{0.49}^{0.49}$ \\
Boo2 & Keck/DEIMOS & \citet{geha_keckdeimos_2026-1} & 17 & 4.25 & 7.65 & Resolved & $1.69_{0.49}^{0.92}$ \\
Boo3 & Keck/DEIMOS & \citet{geha_keckdeimos_2026-1} & 16 & 4.21 & 7.52 & Resolved & $4.96_{1.72}^{2.06}$ \\
CB & Keck/DEIMOS & \citet{geha_keckdeimos_2026-1} & 82 & 4.85 & 111.84 & Resolved & $3.40_{0.50}^{0.61}$ \\
CVn1 & Keck/DEIMOS & \citet{geha_keckdeimos_2026-1} & 254 & 5.08 & 3319.30 & Resolved & $7.56_{0.37}^{0.54}$ \\
CVn2 & Keck/DEIMOS & \citet{geha_keckdeimos_2026-1} & 30 & 4.51 & 180.34 & Resolved & $5.57_{1.13}^{0.95}$ \\
Car1 & Magellan/MMFS,M2FS & \Centerstack{\citet{walker_slar_2009,walker_magellanm2fs_2023}} & 1264 & 5.26 & 82331.55 & Resolved & $6.65_{0.14}^{0.16}$ \\
Car2 & \Centerstack{Magellan/IMACS,AAT/AAOmega+2dF,\\VLT/FLAMES+GIRAFFE} & \citet{li_ships_2018} & 17 & 4.25 & 1000.37 & Resolved & $3.78_{0.73}^{1.00}$ \\
Car3 & \Centerstack{Magellan/IMACS,AAT/AAOmega+2dF,\\VLT/FLAMES+GIRAFFE} & \citet{li_ships_2018} & 5 & 3.39 & 46.48 & Resolved & $4.17_{1.81}^{2.17}$ \\
Cen1 & Magellan/IMACS & \citet{heiger_reading_2024} & 32 & 4.54 & 420.58 & Resolved & $4.00_{0.43}^{0.68}$ \\
Col1 & Keck/DEIMOS & \citet{geha_keckdeimos_2026-1} & 10 & 3.93 & 25.71 & Resolved & $5.12_{1.39}^{2.41}$ \\
Cra2 & AAT/AAOmega+2dF & \citet{ji_kinematics_2021} & 141 & 4.97 & 78.47 & Resolved & $2.27_{0.24}^{0.34}$ \\
Dra & Keck/DEIMOS & \citet{geha_keckdeimos_2026-1} & 999 & 5.24 & 24628.56 & Resolved & $9.56_{0.22}^{0.31}$ \\
Dra2 & Keck/DEIMOS & \citet{geha_keckdeimos_2026-1} & 25 & 4.43 & 0.00 & Unresolved & $<2.65$ \\
Eri2 & VLT/MUSE,Magellan/IMACS & \Centerstack{\citet{zoutendijk_muse-faint_2021}\\\citet{li_farthest_2017}} & 92 & 4.88 & 673.43 & Resolved & $7.69_{0.83}^{1.08}$ \\
Eri4 & Keck/DEIMOS & \citet{geha_keckdeimos_2026-1} & 19 & 4.30 & 69.62 & Resolved & $4.06_{0.65}^{1.26}$ \\
For & Keck/DEIMOS,Magellan/MMFS & \citet{walker_slar_2009} & 3465 & 5.32 & 1939485.61 & Resolved & $12.09_{0.14}^{0.16}$ \\
Gru1 & Magellan/IMACS,M2FS & \Centerstack{\citet{chiti_magellanimacs_2022}\\\citet{walker_magellanm2fs_2016}} & 6 & 3.55 & 16.10 & Resolved & $2.33_{0.89}^{1.24}$ \\
Gru2 & Magellan/IMACS & \citet{simon_birds_2020} & 21 & 4.35 & 1.85 & Unresolved & $<1.85$ \\
Herc & Keck/DEIMOS & \citet{geha_keckdeimos_2026-1} & 43 & 4.65 & 13.07 & Resolved & $2.21_{0.68}^{0.61}$ \\
Hor1 & VLT/FLAMES+GIRAFFE & \citet{koposov_kinematics_2015} & 5 & 3.39 & 136.25 & Resolved & $5.14_{1.43}^{2.70}$ \\
Hya2 & Keck/DEIMOS & \citet{geha_keckdeimos_2026-1} & 12 & 4.05 & 0.00 & Unresolved & $<4.15$ \\
Hyi & Magellan/M2FS & \citet{koposov_snake_2018} & 30 & 4.51 & 400.22 & Resolved & $3.03_{0.39}^{0.58}$ \\
Leo1 & Keck/DEIMOS & \citet{geha_keckdeimos_2026-1} & 795 & 5.22 & 19749.61 & Resolved & $9.23_{0.25}^{0.27}$ \\
Leo2 & Keck/DEIMOS & \citet{geha_keckdeimos_2026-1} & 305 & 5.11 & 4801.95 & Resolved & $7.27_{0.26}^{0.45}$ \\
Leo4 & Keck/DEIMOS & \citet{geha_keckdeimos_2026-1} & 24 & 4.42 & 37.59 & Resolved & $3.26_{0.87}^{0.63}$ \\
Leo5 & Keck/DEIMOS & \citet{geha_keckdeimos_2026-1} & 16 & 4.21 & 21.47 & Resolved & $3.03_{0.96}^{1.06}$ \\
Leo6 & Keck/DEIMOS & \citet{geha_keckdeimos_2026-1} & 9 & 3.86 & 6.81 & Resolved & $3.06_{0.88}^{2.16}$ \\
Peg3 & Keck/DEIMOS & \citet{geha_keckdeimos_2026-1} & 15 & 4.18 & 8.09 & Resolved & $2.67_{0.95}^{1.10}$ \\
Peg4 & Keck/DEIMOS & \citet{geha_keckdeimos_2026-1} & 24 & 4.42 & 41.32 & Resolved & $2.96_{0.65}^{1.16}$ \\
Phe2 & VLT/FLAMES+GIRAFFE & \citet{fritz_gaia_2019} & 5 & 3.39 & 6.02 & Resolved & $9.67_{4.56}^{3.86}$ \\
Pic2 & Magellan/IMACS & \citet{pace_spectroscopic_2025} & 12 & 4.05 & 27.97 & Resolved & $3.62_{1.00}^{1.24}$ \\
Pisc2 & Keck/DEIMOS & \citet{geha_keckdeimos_2026-1} & 12 & 4.05 & 8.16 & Resolved & $3.67_{1.10}^{1.77}$ \\
Ret2 & Magellan/M2FS & \citet{walker_bayesian_2015} & 25 & 4.43 & 97.42 & Resolved & $3.48_{0.60}^{0.69}$ \\
Ret3 & VLT/FLAMES+GIRAFFE & \citet{fritz_gaia_2019} & 3 & 2.85 & -0.00 & Unresolved & $<11.73$ \\
Scl & Keck/DEIMOS, Magellan/MMFS & \citet{walker_slar_2009} & 2681 & 5.31 & 1040141.28 & Resolved & $9.95_{0.13}^{0.14}$ \\
Seg1 & Keck/DEIMOS & \citet{geha_keckdeimos_2026-1} & 60 & 4.76 & 31.39 & Resolved & $4.24_{1.03}^{0.73}$ \\
Seg2 & Keck/DEIMOS & \citet{geha_keckdeimos_2026-1} & 30 & 4.51 & 0.02 & Unresolved & $<2.18$ \\
Sext & Keck/DEIMOS, Magellan/MMFS & \citet{walker_slar_2009} & 745 & 5.22 & 79667.73 & Resolved & $8.24_{0.24}^{0.22}$ \\
Tri2 & Keck/DEIMOS & \citet{geha_keckdeimos_2026-1} & 7 & 3.68 & -0.00 & Unresolved & $<4.05$ \\
Tuc2 & Magellan/IMACS,MIKE,MagE,M2FS & \citet{chiti_detailed_2023} & 18 & 4.28 & 211.12 & Resolved & $3.64_{0.94}^{0.53}$ \\
Tuc3 & Magellan/IMACS & \citet{simon_nearest_2017} & 26 & 4.45 & 2.73 & Unresolved & $<1.73$ \\
Tuc4 & Magellan/IMACS & \citet{simon_birds_2020} & 9 & 3.86 & 129.72 & Resolved & $4.48_{1.04}^{1.60}$ \\
Tuc5 & Magellan/MIKE,IMACS & \citet{hansen_chemical_2024} & 4 & 3.17 & 2.06 & Unresolved & $<4.19$ \\
UMa1 & Keck/DEIMOS & \citet{geha_keckdeimos_2026-1} & 37 & 4.59 & 285.48 & Resolved & $6.91_{0.91}^{1.35}$ \\
UMa2 & Keck/DEIMOS & \citet{geha_keckdeimos_2026-1} & 64 & 4.78 & 151.41 & Resolved & $6.03_{0.88}^{1.21}$ \\
UMi & Keck/DEIMOS & \citet{geha_keckdeimos_2026-1} & 832 & 5.23 & 16155.03 & Resolved & $8.92_{0.25}^{0.26}$ \\
W1 & Keck/DEIMOS & \citet{geha_keckdeimos_2026-1} & 53 & 4.72 & 264.78 & Resolved & $4.74_{0.65}^{0.84}$ \\
    \end{tabular}
    \label{tab:ufds}
\end{table*}

%\newpage

\bibliography{main}{}
\bibliographystyle{aasjournal}

\end{document}